\documentclass[aps,prd,groupedaddress,nofootinbib]{revtex4}
\usepackage{graphicx}
\usepackage{amssymb,amsfonts}
\begin{document}
\def\ie{{\it i.e.}}
\def\etl{{\it et al.}}
\def\Rmo{{\it R}-mode }
\def\Rmos{{\it R}-modes }
\def\rmo{{\it r}-mode }
\def\rmos{{\it r}-modes }
\def\R{{\mathbb R}}
\def\ov{\over}
\def\l({\left(}
\def\r){\right)}
\def\la{\left[}
\def\ra{\right]}
\def\bc{\begin{center}}
\def\ec{\end{center}}
\def\be{\begin{equation}}
\def\ee{\end{equation}}
\def\bi{\begin{itemize}}
\def\ei{\end{itemize}}
\def\dt{\delta}
\def\Dt{\Delta}
\def\alf{\alpha}
\def\gam{\gamma}
\def\th{\vartheta}
\def\ph{\varphi}
\def\pt{\partial}
\def\eps{\varepsilon}
\def\omg{\Omega}
\def\nb{\nabla}
\def\dv{\rm{div}}
\def\d{\rm{d}}
\def\oa{\overrightarrow}
\def\vnb{\vec{\nabla}}


\title{Inertial modes in slowly rotating stars : an evolutionary description}

\author{Lo\"\i c Villain}
\email[]{loic.villain@obspm.fr}
\author{Silvano Bonazzola}
\email[]{bona@mesiob.obspm.fr}

\affiliation{Laboratoire Univers et TH\'eories (F.R.E. 2462 du CNRS)
/ Observatoire de Paris, F-92195 Meudon Cedex, France}

\date{\today}
\begin{abstract}
We present a new hydro code based on spectral methods using spherical
coordinates. The first version of this code aims at studying time
evolution of inertial modes in slowly rotating neutron stars. In this
article, we introduce the anelastic approximation, developed in
atmospheric physics, using the mass conservation equation to discard
acoustic waves. We describe our algorithms and some tests of the
linear version of the code, and also some preliminary linear
results. We show, in the Newtonian framework with differentially
rotating background, as in the relativistic case with the strong
Cowling approximation, that the main part of the velocity quickly
concentrates near the equator of the star. Thus, our time evolution
approach gives results analogous to those obtained by Karino {\it et
al.} \cite{karino01} within a calculation of
eigenvectors. Furthermore, in agreement with the work of Lockitch {\it
et al.} \cite{lockandf01}, we found that the velocity seems to always
get a non-vanishing polar part.
\end{abstract}


\maketitle

\section{Introduction}
In 1998, Andersson \cite{anders98} discovered that in any relativistic
rotating perfect fluid, {\it r}-modes\footnote{In this paper, we shall
use the terminology of the hydro community : inertial modes are the
oscillatory modes of a fluid whose restoring force is the Coriolis
force. \Rmos [or purely toroidal (axial) modes] then form a sub-class
of inertial modes.  The later were discovered by Thompson in 1880
\cite{thom80}. The reader can find a short point of history in
Rieutord 2001 \cite{rieut01}.}  are unstable via the coupling with
gravitational radiation. This is just a particular case of the
\mbox{Chandrasekhar-Friedman-Schutz} (CFS) mechanism discovered in
1970 (see \cite{chandra70} and \cite{friedschu78}).  However, a
characteristic of \rmos is the fact that this instability is expected
whatever the speed of rotation of the star (see also Friedman {\it et
al.} 1998 \cite{friedmor98}). This discovery triggered an important
activity in the neutron stars (NS) community since such inertial
instabilities might contribute to explanation of the observed relative
slow rotation rates of young and recycled NS (see Bildsten
\cite{bil98}). Moreover, such an instability could in principle make
of NS efficient sources of gravitational waves (GW) for observation by
the ground interferometric detectors under construction or their
direct descendants. Nevertheless, as various not always well-known
physical processes take place which might inhibit the growth of {\it
r}-modes in real NS, their relevance is still questionable at this
time. More details can be found in the reviews by Andersson and
Kokkotas 2001 \cite{andko01} and Friedman and Lockitch 2001
\cite{friedlock01}.\\

The present study was originally motivated by the results announced by
Lindblom {\it et al.} in 2001 \cite{lintv01}. These authors computed, in
the highly non-linear regime, the evolutionary track of the CFS instability of
\rmos in a Newtonian NS with a magnified approximate post-Newtonian
radiation reaction (RR) force. In their calculation, they found a rapid growth
of the mode, until strong shocks appeared and quickly damped it. A recent
article by Arras {\it et al.} \cite{arr02} asserts that this phenomenon was
just an artifact linked with the huge RR force. Yet, we found those results so
interesting and so surprising that we decided to try to reproduce 
and better understand them with a different approach. The code of Lindblom
{\it et al.} \cite{lintv01} was written in cylindrical coordinates with 3D
finite difference scheme and they studied fast rotating stars. We chose to
create a spectral code in spherical coordinates and to begin with a slowly
rotating NS.\\

 The use of spectral methods was motivated by the fact that they can provide a
deep insight in describing the turbulence generated by quadratic terms. This is
analogous to what is done with Fourier analysis in the study of homogeneous
turbulence (e.g. Lesieur 1987 \cite{lesieur87}). Furthermore, we chose a slowly
rotating NS for several reasons that will be explained in the following.\\

Since NS are known for their quite rapid rotation and since inertial
modes have for restoring force the Coriolis force, an {\it a priori}
reasoning would probably lead to the conclusion that the only - or
most - interesting case of inertial modes to study is the case of
modes in a fast rotating NS. However, the final answer is not so easy
to decide, as among observational data and among models of cooling of
pulsars, many things in fact support slowly rotating newborn single
NS. For instance, the estimations of the rotation periods at birth of
the two historical X-rays and radio pulsars whose ages are known
exactly (the 820 year old 65.8 ms period pulsar in SNR 3C58
\cite{MSPS02}, \cite{CS2002} and the 948 year old 33 ms period Crab
pulsar) are respectively 60 ms \cite{MSPS02} and 14 ms
\cite{SBJM94}. Moreover, compared with assessment of the angular
momentum of isolated supergiant stars (whose core collapses will give
type II supernov\ae{} and then potentially NS), these numbers show
that important losses of angular momentum are expected to happen
during the stellar evolution. Furthermore, this is the
same concerning the giant phase of less massive stars that give white
dwarfs. Indeed, the observed rotation periods of 19 white dwarfs range
between 12.1 min and 12 days with a median value of about 1 hour
\cite{Scmi01} but, if the sun shrank without losing any angular
momentum into a white dwarf of the same mass with a radius of 5000 km,
its rotation period would be of a few minutes only. In the stellar
evolution community, several ways to explain these weak angular
momenta can be found and the final answer is still not clear. But a
quite admitted idea is that magnetic field plays probably the key-role via
magnetic braking type mechanisms (Schatzman 1962 \cite{scha62}). And
whatever the actual mechanism, the current conclusion in this community
is that a huge amount of angular momentum is supposed to be lost
during the stellar evolution for main sequence stars of any mass. Then,
the most difficult thing to explain in stellar evolution physics seems
to be the reason why these rotation periods at birth (of the orders of
1 hour for white dwarfs and of 1 ms for NS) are so {\bf small}
(e.g. Spruit 1998 \cite{spr98} and Spruit and Phinney 1998
\cite{spph98}): stellar evolution scenarios do not expect of baby NS
to be fast rotating.\\

 However, it can be mentioned that some isolated fast rotating NS are found.
But, for the most part of them, the current idea is that these stars have been
recycled in binary systems. In these systems, where fast rotating NS exist,
the NS is thought to have been spun up by the accreted matter. The recent
discoveries of accreting millisecond pulsars (XTEJ0929-314, SAXJ1808.4-3685,
XTEJ1751-305 and 4U 1636-53) whose rotation frequencies are respectively
185 Hz, 401 Hz, 435 Hz and 581 Hz \cite{STRO02} (corresponding to respective
periods of 5.4 ms, 2.5 ms, 2.3 ms and 1.7 ms) support the above scenario. Yet,
there is also an example of single ms pulsar associated to a supernova remnant:
PSR J0537-6910 whose period is 16 ms, whose age is about 5000 yr and which
is supposed to have had an initial period of a few ms (see for instance
\cite{mar98}). Hence, in spite all the above arguments, the existence
of rapidly rotating baby NS should not be completely excluded. Moreover, a
fast rotating baby NS with very weak magnetic field and consequently very weak
losses of angular momentum due to magnetic braking like mechanisms during the
supergiant phase would escape to the observations as pulsars but would be very
interesting as GW sources.\\

Nevertheless, such a discussion does not make clear the important
point. Indeed, whatever the value of the frequency, it does not
directly tell if a pulsar is fast rotating or if it is not. What says,
in a particular work, if a NS can be regarded as (quite) slowly
rotating is the relative importance of the deformation for this
study. In a more general framework, this question is settled by the
ratio between the angular velocity of the pulsar and its Keplerian
angular velocity. Even for the fastest rotating known pulsars (that
are in binary systems) this ratio is less than a third, since the
Kepler frequency is around 1 ms (see \cite{haen89}). Furthermore, what
is implied in the hydrodynamics of a NS is not this ratio, but its
square. Hence, in appropriate units, this factor is less than ten per
cent of the {\it Coriolis force} for most of the NS. Anyway, there is
a last crucial issue. In a Newtonian star, even if this factor is a
few per cent correction in the equation of motion, it is fundamental
since it creates a coupling between polar and axial part of the velocity.
Yet, in a relativistic star, the situation is completely different as
noticed Kojima \cite{koj98}. Indeed, in a relativistic rotating star,
the frame dragging term has the same qualitative result: it
makes a coupling between the polar and the axial parts of the
velocity.  But, in appropriate units, this coupling is scaled by the
ratio between the angular velocity and the Keplerian angular velocity
and not by the square of this ratio. Hence, even for the fastest
rotating known NS, the deformation introduces a kind of second order
correction that can be neglected in a first approach\footnote{As an
example, we verified that, for a rotation frequency of 300 Hz in a
star of $1.7\,M_\odot$ with a fully relativistic code for stationary
configuration of rotating stars, the coupling between the spherical
harmonic terms due to the drag effect is one order of magnitude larger
than the coupling due to the deformation of the star.}.\\

 Thus, our choice of using the slow rotation limit as a first step was
motivated by all the astrophysical reasons mentioned above: even for
ms pulsar, the slow rotation approximation is still quite good. But,
secondly, we wanted to look for a possible saturation due to
non-linear coupling that may occur before a highly non-linear regime
is reached. To better understand such a phenomenon, we thought it was
probably wiser to begin with an easier situation in which there are
not several effects with consequences of the same orders of
magnitude. Thus, we began to build a non-linear hydrodynamics code
using the Newtonian theory of gravity and the slow rotation
approximation. But, once a linear Newtonian code had been written
(first step to a non-linear version), upgrading it to a general
relativity (GR) linear code with {\it strong Cowling
approximation}\footnote{In December 2001, during a workshop on \rmos
which took place at the Meudon site of Paris Observatory, Carter
suggested to call {\it strong} Cowling approximation the approximation
in which all the coefficients of the metric are frozen. This name was
chosen to contrast with what should be called the {\it weak} Cowling
approximations where some perturbations of the metric are allowed. See
for instance the work of Ruoff {\it et al.} \cite{ruoff01},
\cite{ruoff01b} and Lockitch {\it et al.}  \cite{lockandf01} using the
Kojima equations \cite{koj92}.}  was quite obvious and we chose it for
a second approach to inertial instabilities. Indeed, as it was already
mentioned above, Kojima first noticed in 1998 \cite{koj98} that the
frame dragging phenomenon makes the relativistic \rmos quite different
from the Newtonian {\it r}-modes.  But concerning this linear
relativistic study, we also decided to begin with the slow rotation
limit to try to get a better understanding of the spherical
relativistic case, before putting what can be seen as a second order
correction.\\

 This article is organized as follows: in Section \ref{dim}, we start by
describing the physical conditions we chose in the Newtonian case. Then we
write the Navier Stokes equations (NSE) in dimensionless form, and the RR force
we introduced in the linear study. This enables us to calculate the
associated characteristic numbers and to observe that the corresponding
numerical problem is a stiff one with typical time scales being orders of
magnitude different. We discuss some approximations that can be
made in Section \ref{approx}. The most important of them is the anelastic
approximation. Section \ref{calib} explains the basic tests of the code, with
for instance the linear $l=m$  \rmos in a rigidly rotating spherical fluid.
This admits an easy and analytical solution of Euler equation (EE) in the
divergence free approximation. We demonstrate that, thanks to spectral
methods, the velocity profile is exactly determined and preserved within the
round-off errors. We also show that the error in the conservation of the
energy is only due to the time discretization and that it vanishes as the
cube of the time-step $\Dt t$. Finally, this section ends with tests of the
anelastic approximation and of the RR force we adopted. Section \ref{difrot}
deals with results obtained in a differentially rotating NS. As expected,
pure \rmos are replaced with inertial modes which are still CFS unstable, even
if they are no longer purely axial. Moreover, we show that if the amplitude of
the difference of angular velocity between the interior and the surface of the
star is sufficiently large, these modes quite rapidly develop a kind of
``singularity'' in the first radial derivative of the velocity. There is then
a kind of {\it concentration of the motion} near the surface and mainly near
the equatorial plane. Our time evolution approach thus provides results of the
same kind of those obtained by Karino {\it et al.} in 2001 \cite{karino01} with
a modes calculation. A viscous term has to be added in order to
regularize the solution, but it does not change the qualitative result. We
shall discuss this result in the conclusion. In Section \ref{grcas}, we give
some first results for the GR case in a slowly rotating NS with
the {\it strong} Cowling approximation. In this framework, the above
conclusions still hold. More results in GR will be given in a following
article in preparation. The Section \ref{conc} contains the conclusion and
discussion. A detailed description of the numerical algorithm is given in the
appendix.

\section{Equations and numbers} \label{dim}

\subsection{The barotropic case in Newtonian gravity} \label{newtnl}

Cooling calculations (e.g. Nomoto and Tsuruta 1987 \cite{nom87}, Yakovlev
{\it et al.} 2001 \cite{yak01}) showed that several minutes are enough to
enable the NS matter to fall far below its Fermi temperature, that is
roughly $10^{10}$ K. Furthermore, in the Newtonian non-linear hydrodynamics
of a not too young slowly rotating NS, it is not worth trying to use a very
sophisticated EOS. We shall therefore adopt a barotropic EOS. With those
assumptions, we have
\be \label{eos}
P=P[n]\text{,}
\ee
where $P$ is the pressure and $n$ the mass density, and the Poisson
equation for the gravitational potential $U$ is
\be \label{pois}
\Dt U = 4 \pi G\, n.
\ee

The Newtonian Navier-Stokes equations (written in the inertial frame
for a rigidly rotating NS) are
\begin{eqnarray}
& \label{NSE}
\l(\pt_t + \omg\, \pt_{\ph}\r)\,\vec{W}+\l(\vec{W} \cdot \vnb\r) \vec{W}
+\,2 \vec{\omg} \wedge \vec{W} + \vnb H = \\
& \frac{1}{n} \, \l( \vnb\l(\vnb \eta \cdot \vec{W}\r) + \vnb\wedge
\l(\vec{W}\wedge \vnb{\eta}\r) + \eta \Dt \vec{W} - \vec{W} \Dt \eta +
\vnb \l(\l(\zeta+ {\eta \over 3}\r) \vnb \cdot \vec{W}\r) \r) +\vec{A}.
\nonumber
\end{eqnarray}

Here we take $(r,\th,\ph)$ for a spherical system of coordinates in
the inertial frame. In this system, $\th=0$ coincides with the
direction of the rotation axis of the NS, which is parallel to the
angular velocity: $\vec{\omg} = \omg\,\vec{e}_z$. $\vec{W}$ is the
velocity {\it in the corotating frame}, {\it i.e.} the part that is
added to the velocity of the rigidly rotating background when a {\it
mode} is present. Note that both velocities can be of the same order
in the non-linear case. This is the reason why we shall not use the
term {\it Eulerian perturbation} for $\vec{W}$. Otherwise, $H= \int
{\frac{1}{n}\frac{{\d} P}{{\d} n} {\d} n}$ is the enthalpy, $\eta$ and
$\zeta$ are respectively the dynamical shear and bulk viscosity
coefficients, and finally $\vec{A}$ contains any external {\it
accelerations} (or force {\it per unit of mass}). The modifications
that are needed in the case of the linear study with differential rotation
or in the relativistic linear case are respectively discussed in sections
\ref{difrot} and \ref{grcas}.\\

In what follows, $\vec{A}$ is mainly the effective gravitational acceleration,
{\it i.e.} the gradient of the difference between the centrifugal potential
$\frac{1}{2} \omg^2 \rho^2$ and the gravitational potential $U$, where $\rho$
is the distance from the rotation axis. In the linear regime, we
will sometimes introduce a RR acceleration ({\it cf.} Blanchet 1993 and 1997
\cite{blan937}) of the form
\[ {A_{RR}}_j\widehat{=}\frac{4}{c^2}v^i\pt_j\Upsilon_i\,\,, \]
where
\be \label{radreac}
\Upsilon_i\widehat{=}\frac{4}{45}\frac{G}{c^5}\epsilon_{ijk}x^jx^m
{{S^{km}}}^{\l(5\r)}[t]
\ee
and
\[ {S_{km}}[t] \widehat{=}\,\epsilon_{pq\l(k\right.} \int {\d}^3 x \,n\,
x_{m\left.\r)}x^pv^q\,\,,\] with
$\l(km\r)\widehat{=}\frac{1}{2}(km+mk)$, $v^i$ being the full velocity
and the superscript $^{\l(5\r)}$ the fifth time derivative, which can
not be easily calculated in a numerical work (see Rezzolla {\it et
al.}  1999 \cite{rezz99}). Note that this formula is valid only if
written with Cartesian components of the tensors and this is the
reason why we did not really make distinction between contra and
covariant components. Finally, we insist on the fact that such an
acceleration does not include any mass multipole but only the current
quadrupole that is the most important coefficient for the emission of
GW by axial modes in a slowly rotating NS.\\

To fulfill the system of equations, we need to add the mass continuity
equation:
\be \label{mascons}
\l(\pt_t + \omg\, \pt_{\ph}\r)\,n + \vec{W} \cdot \vnb n + n\, \vnb \cdot 
\vec{W} = 0.
\ee

Taking into account that the EOS is a barotropic one, the preceding equation
can be written in a slightly different way, but we will discuss it in the
Section \ref{approx}.\\

As far as inertial modes are concerned, the typical time scale and length scale
are $2\pi\omg^{-1}$ and $R$, characteristic length of the background star. The
velocity associated with those values, $R\,\omg$, can be very far from the
characteristic velocity of the mode. Another velocity, scaling that of the
mode, must be introduced. But instead, we introduce $\alf$, a pure number
that is defined as the ratio of these two velocities. Otherwise, the typical
mass is obviously the star's, $M_{\star}\sim 1.4 M_{\odot}$.
Bearing all this in mind, we define the following dimensionless variables:
$$
\xi\widehat{=}{1 \over R}r, \hspace{1cm} \tau\widehat{=}t\,\omg, \hspace{1cm}
\vec{V}\widehat{=}{1 \over \alf\,\omg\,R}\vec{W},
$$
\be
\label{var}
d\la\vec{\xi}\ra\widehat{=}{1 \over n_0} n\la\vec{r}\ra,\hspace{1cm}
s\la\vec{\xi}\ra\widehat{=}{1 \over \nu\,n_0\,\omg\,R^2}\eta\la\vec{r}\ra,
\ee
$$
b\la\vec{\xi}\ra\widehat{=}\frac{1}{\lambda\,n_0\,\omg\,R^2}\zeta\la\vec{r}\ra,
\hspace{1cm}\vec{e}_z\widehat{=}\frac{1}{\omg}\vec{\omg},
$$
$$
\tilde{H}\la\vec{\xi}\ra\widehat{=}\frac{1}{\alf\,\omg^2\,R^2}H\la\vec{r}\ra
\hspace{1cm}\textrm{and}\hspace{1cm}
\vec{\Sigma}\la\vec{\xi}\ra\widehat{=}
\frac{1}{\alf\,\omg^2\,R}\vec{A}\la\vec{r}\ra.
$$
Thus the motion equations are written [with same conventions than in
Eq. (\ref{NSE})] as
\begin{eqnarray}
& \label{NSED}
\l(\pt_{\tau} + \pt_{\ph}\r)\,\vec{V}+\,\alf\,\l(\vec{V} \cdot \vnb\r) \vec{V}
+\,2  \vec{e}_z \wedge \vec{V} + \vnb \tilde{H} = \nonumber \\
& \frac{\nu}{d} \, \l( \vnb\l(\vnb s \cdot \vec{V}\r) + 
\vnb\wedge\l(\vec{V}\wedge\vnb{s}\r) + \,s \Dt \vec{V}- \vec{V} \Dt \,s
+ \vnb \l(\l(\frac{\lambda}{\nu}\,b+\frac{s}{3}\r) \vnb\cdot\vec{V}\r) \r)
+\vec{\Sigma}, \nonumber
\end{eqnarray}
where the $\vec{\nb}$ and $\Dt$ operators are now performed with dimensionless
variables ($\xi, \th, \ph$) instead of ($r, \th, \ph$).\\

Here, we have introduced the following notations:
\bi
\item [-] $\alf$ a pure number, characteristic of the initial amplitude of the
mode, but also of the ratio of the non-linear term and the Coriolis force. With
our conventions, {\it $\alf$ is twice the usual Rossby number} and $\vec{V}$
is a vector whose norm at $t=0$ is equal to 1 in a given point
on the equator;
\item [-] $n_0 = M_{\star}/(\frac{4}{3} \pi R^3)$. For the full slow
rotation limit approximation (only terms linear in $\omg$ and spherical shape),
$R$ would be the radius of the star and then $n_0$ the mean density.
\item [-] $s$(hear) and $b$(ulk) dimensionless functions, and $\nu$ and
$\lambda$ pure numbers, all chosen in such a way that if there is any
viscosity, $s$ and $b$ are equal to $1$ in a specific position. Typically,
for a single fluid model, this would be the center of the star. Nevertheless,
it is worth pointing out that to build a more realistic model of NS, several
different layers could be made to coexist. In this case, there would be
different $\nu$ and $\lambda$, hugely depending on the shell (see for instance
Haensel {\it et al.} 2001 \cite{haen01}). With our definitions, {\it those
numbers, $\nu$ and $\lambda$, are twice the usual Ekman numbers}, and they then
quantify the ratios between the viscosities and the Coriolis force.
\item [-] $\tilde{H}$ and $\vec{\Sigma}$ respectively the dimensionless
enthalpy and the dimensionless external accelerations, both scaled by the
inverse of $\alf$.
\ei

Concerning the latter quantities, we cut them in two parts and use for
variables the difference between their present values and their values in the
steady state. For the background parts, we then have
\be
\vec{\Sigma}_0 = -\vec{\nb} \tilde{U} + \vec{\nb} \frac{\l(\xi \sin \th\r)^2}
{2\,\alf}
\hspace{0.5cm}
\text{with}
\hspace{0.5cm}
\vec{\nb} \tilde{H}_0 = \vec{\Sigma}_0.
\ee

This equality enables the background to be solution of the NSE and can
be solved separately. From now until the end, we consider that this condition
is realized, and we forget about $\vec{\Sigma}_0$ and $\tilde{H}_0$ (except in
the mass conservation equation where the latter still appears as an
{\it external} parameter).\\

Finally, some words about the Newtonian gravitational field. Inertial modes are
current oscillations associated with small density variations. In this context,
it is natural to use the Cowling approximation (Cowling 1941 \cite{cow41})
which consists in forgetting fluctuations of the gravitational field (see the
relevance of this approximation for Newtonian \rmos in Saio
1982 \cite{saio82}). We will apply it for the non-linear study. Nevertheless,
for the Newtonian linear study, we shall see later that depending on the way
mass conservation is treated, it may be not necessary (see Section \ref{approx}
for more details). In this case, we have
\be
\vec{\Sigma} = \vec{\Sigma}_0 + \vec{\sigma}
\hspace{0.5cm}
\text{with}
\hspace{0.5cm}
\vec{\sigma} = - \vnb{\dt \tilde{U}}
\ee
where $\dt \tilde{U}$ is the Eulerian perturbation of the dimensionless
gravitational potential.

\subsection{Dimensionless equations of motion for the spherical components of
the velocity}

The equations of motion for the spherical components of the velocity in the
orthonormal basis associated with $(\xi, \th, \ph)$ are given by
\begin{eqnarray}
& \pt_{\tau} V_r + \pt_{\ph} V_r + \l(\vec{V}\cdot\vnb\r)V_r - \frac{1}{\xi}
\l(V_{\th}^2 + V_{\ph}^2\r) - 2\, \sin \th\, V_{\ph} + \pt_{\xi}
\tilde{h} =\nonumber\\
& \frac{\nu}{d} \, \l( \l(\pt_{\xi} s\r)\l(2 \pt_{\xi} V_r\r) + s \l(\Dt 
V_r - \frac{2}{\xi^2} \l(\frac{1}{\sin \th} \pt_{\th}\l(\sin \th\,V_{\th}\r) +
\frac{1}{\sin \th}\pt_{\ph}V_{\ph} + V_r\r)\r)\right. \nonumber\\
&\left. - \pt_{\xi} s \l(\vnb\cdot\vec{V}\r) +
\pt_{\xi}\l(\l(\frac{\lambda}{\nu}\,b+
\frac{s}{3}\r)\vnb\cdot\vec{V}\r) \r) + \sigma_r, \nonumber
\end{eqnarray}
\begin{eqnarray}
& \label{NSEDC}
\pt_{\tau} V_{\th} + \pt_{\ph} V_{\th} + \l(\vec{V}\cdot\vnb\r)V_{\th} +
\frac{1}{\xi}\l(V_{\th}V_r - V_{\ph}^2 \cot \th\r) - 2\,\cos \th\,V_{\ph} + 
\frac{1}{\xi} \pt_{\th}\tilde{h} = \nonumber\\
& \frac{\nu}{d} \, \l( \l(\pt_{\xi} s\r)\l(\pt_{\xi} V_{\th} +
\frac{1}{\xi}\l(\pt_{\th} V_r - V_{\th}\r)\r) + s \l(\Dt V_{\th} +
\frac{1}{\xi^2} \l(2 \pt_{\th}V_r
- \frac{V_{\th}}{\sin^2[\th]} -\frac{2 \cos \th }{\sin^2[\th]}\pt_{\ph}
V_{\ph}\r)\r) \right.\\
&\left. + \frac{1}{\xi}\pt_{\th} \l(\l(\frac{\lambda\,b}{\nu}+\frac{s}{3}\r)
\vnb\cdot\vec{V}\r) \r) + \sigma_{\th}, \nonumber
\end{eqnarray}
\begin{eqnarray}
& \pt_{\tau} V_{\ph} + \pt_{\ph} V_{\ph} + \l(\vec{V}\cdot\vnb\r)V_{\ph} +
\frac{1}{\xi}\l(V_{\ph}V_r + V_{\ph}V_{\th} \cot[\th]\r) + 2 \l(\sin
\th\,V_r +
\cos \th\,V_{\th}\r) + \frac{1}{\xi \sin \th} \pt_{\ph}\tilde{h} = \nonumber\\
& \frac{\nu}{d} \, \l( \l(\pt_{\xi} s\r)\l(\pt_{\xi} V_{\ph} + \frac{1}
{\xi \sin \th}\l(\pt_{\ph} V_r - \sin \th\,V_{\ph}\r)\r) + s \l(\Dt V_{\ph} +
\frac{1}{\xi^2}\l(\frac{2}{\sin \th} \pt_{\ph}V_r + \frac{2 \cos \th }
{\sin^2[\th]}\pt_{\ph}V_{\th}- \frac{V_{\ph}}{\sin^2[\th]}\r)\r)
\right. \nonumber \\
&\left. + \frac{1}{\xi \sin \th}\pt_{\ph}\l(\l(\frac{\lambda\,b}{\nu}+
\frac{s}{3}\r)\vnb\cdot\vec{V}\r) \r) + \sigma_{\ph},\nonumber
\end{eqnarray}
where
\[
\Dt f \widehat{=} \frac{1}{\xi} \pt_{\xi}^2\l(\xi\,f\r)
+ \frac{1}{\xi^2 \sin \th} \pt_\th\l(\sin \th\,\pt_\th f\r)
+ \frac{1}{\xi^2 \sin^2 \th} \pt_{\ph}^2 f
\]
is the scalar Laplacian, while
\[\l(\vec{V}\cdot\vnb\r) f \widehat{=} V_r \pt_\xi f + V_\th \frac{1}{\xi}
\pt_\th f + \frac{1}{\xi \sin \th} V_\ph \pt_\ph f
\]
and
\[
\vnb\cdot\vec{V} \widehat{=} \frac{1}{\xi^2} \pt_\xi\l(\xi^2\,V_r\r) + \frac{1}
{\xi\,\sin \th} \pt_\th\l(\sin \th\, V_\th\r) + \frac{1}{\xi\,\sin \th}\pt_\ph
V_\ph.
\]

In the above equations, we assume $s$ to only depend on $\xi$ in the
$(\xi, \th, \ph)$ system of coordinates. This assumption is linked
with the slow rotation limit we should apply from now. This
approximation supposes that the star is slowly rotating compared with
its Kepler frequency. As explained in the introduction, observational
data make this assumption credible, while known pulsars seem to rotate
with a velocity smaller than a third of their Kepler velocity. In the
lowest framework, only terms linear in $\omg$ are kept, and the
background star is supposed to have preserved its spherical
shape. Yet, to improve the slow rotation approximation, one can go a
step further and take into account the deformation of the star,
assuming it is a small order effect (or small quantum number $l$
effect). To make this improvement, it is sufficient to introduce a new
variable $\chi[\xi,\th]$ that is equal to the unity at the surface,
that coincides with isosurfaces of pressure, enthalpy and density, and
that decomposes on $m=0$ Legendre functions\footnote{Restricting to
the case $m=0$ is sufficient for an isolated NS, but in a binary
system, $m=2$ should also be included.}. With our algorithm, the
easiest way to proceed would be to keep the spherical basis for the
vectors, and to express all spatial operators depending on
($\xi,\th,\ph$) as the sum of the new operators depending on
($\chi,\th,\ph$) and of terms to interpret as {\it applied
accelerations}. More details can be found in Bonazzola {\it et al.}
1997, 98, 99 \cite{bonagm97}, \cite{bonagm98} and
\cite{bonagm99}. This will not be done in this article, for reasons
that were also explained in the introduction. Finally, note that we do
not really solve these equations. Indeed, we use the Helmholtz theorem
(see for instance Morse and Feshbach 1953 \cite{mors53}) that says
that any vector of $\R^3$ can be in an unique way written as the sum
of a divergence free vector and of the gradient of a potential, given
its normal component over the boundary:
\be
\vec{V} = \vnb{\psi} + \vnb \wedge \vec{{\cal V}}.
\ee

 Instead of working with the above equations [Eq.(\ref{NSEDC})], we use the
equation on the scalar potential $\psi$ and the equations on the $\th$ and
$\ph$ components of the divergence free vector $\vnb \wedge \vec{{\cal V}}$.
Most of the time, these equations cannot be reached analytically. The way to
proceed is then to separate the potential and the divergence free parts
of the initial equations by numerically solving Poisson like equations
obtained by taking their divergence. The reader can find in the appendix
more details about these algorithms.

\subsection{Characteristic numbers} \label{charnumb}

To better understand what are the dominant processes in the dynamics
of NS, it is worth estimating characteristic time scales and
associated numbers.  The acoustic time, {\it i.e.} the duration of the
acoustic waves travel across the NS is by far the shortest. For a
typical NS, it is about $2.10^{-4}$ seconds \footnote{This time is of
the same order of magnitude as the inverse of the frequency of the
fundamental pressure mode, the so-called {\it p}-mode.}.  As we are
dealing with slowly rotating NS, the period of rotation should be more
than $\sim 2.10^{-3}$ seconds ($\equiv 500 Hz$). It follows that this
time is at least one order of magnitude greater than the acoustic
time. It is then the same for the typical period of inertial modes,
their frequency being (at linear order) proportional to $\omg$.\\

Another time scale is the viscous damping time associated with viscosity:
\[
T_v \sim \frac{n_0\,R^2}{\max[\eta,\zeta]}
\]
or
\[
T_v \sim \frac{1}{2\,\max \la E_s,E_b \ra\,\omg}
\]
where $E_s$ and $E_b$ are respectively the Ekman number for shear and bulk
viscosity. In other words, the Ekman number must be interpreted as
characteristic of the ratio between the period and the viscous time. This
number is typically less than $10^{-7}$ and the viscous time is more than 7
orders of magnitude greater than the period (see Cutler {\it et al.} 1987
\cite{cut87} for more precise values). A flow will be said ``rotation
dominated'' if the above Ekman and Rossby numbers are small compared to the
unity. Note that another usual hydrodynamical number appears with them: the
Reynolds number, prodrome of turbulence. In a rotating fluid, it is defined by
the ratio between Rossby and Ekman numbers, or in any fluid by the ratio
between the non-linear and the viscous terms.\\

The last typical time to evaluate here is the instability rising time $T_g$
associated with the RR force. To get an idea of its value, we come back to
the dimensionless RR acceleration, formulae (\ref{radreac}) and
definitions (\ref{var}). Analytical calculation with $V^i$ being
the $l=m$ linear \rmo(with time dependent amplitude)
\be
\vec{V} = {\cal A}[\tau] \frac{1}{m} \Re \la {\xi}^m\,\vec{\xi}
\wedge \vnb Y_{mm} \ra
\ee

or in the spherical orthonormal basis
\be \label{rmodequa}
\vec{V}={\cal A}[\tau] \,\left|\begin{array}{l}
0\\
{\xi}^m (\sin \th)^{m-1}\, \sin[m \ph]\\
{\xi}^m (\sin \th)^{m-1}\, \cos[\th]\,\cos[m \ph]\\
\end{array}
\right.
\ee
(where $\Re$ means the real part of the complex function), gives in dimensioned
variables at the lowest order for the $m=2$ {\it r}-modes
\be \label{tg}
\pt_t {\cal A}[t] = \frac{G R^7 \omg^6 n_0}{c^7}
\frac{2^{16}\,\pi}{3^8\,5^2\,7} {\cal A}[t] \widehat{=} \frac{1}{T_g}
{\cal A}[t].
\ee

For a typical spherical NS with $R=10$ km, $M=1.4\,M_\odot$ and $\omg \sim
(2\,\pi) 200$ Hz, $T_g$ is something like $10^8$ periods of the NS. Thus,
depending on the viscosity given by the EOS, it will or will not be larger
than $T_v$, and then, the inertial instability will or will not be relevant.
At this step, a new typical number seems natural to introduce. We shall
propose to call ``Chandra number''\footnote{Chandrasekhar was the first who
studied the gravitational radiation driven instability for the $l=m=2$
fundamental modes of uniform density MacLaurin spheroid in 1970. See
\cite{chandra70}.} the ratio between the viscous time $T_v$ and the rising
time $T_g$. In the same spirit as the Rossby and Ekman numbers, it should
also quantify the ratio between the viscous and RR forces:
\be \label{chand}
Ch = \frac{2^{16}\,\pi}{3^8\,5^2\,7} \frac{G\,{n_0}^2\,R^9\,\omg^6}
{c^7\,\max[\eta,\zeta]}.
\ee

Note finally that with this factor ahead of the physical parameters, we ensure
the bifurcation value of $Ch$ to be of the order of the unity, at least for
the linear $l=m=2$ {\it r-}mode. Indeed, this point is easily illustrated by
looking at NSE for the linear $l=m=2$ \rmo with time dependent amplitude and a
shear viscosity of the form $s=1-{\xi}^2$. This viscosity that vanishes at the
surface implies no need to add more boundary conditions (BC) and gives the
exact (at the lowest order for the RR force) differential equation for the
amplitude:
\[ \pt_t {\cal A}[t] = \frac{1}{T_g} {\cal A}[t]
\l(1 - \frac{2}{Ch} \r)
\]
where one easily sees that with this shape, the viscosity wins the battle
against RR force for $Ch < 2$. All the previous numbers are gathered in
Table \ref{numbs}.\\

To end with this short discussion, we should insist on the most important
conclusion that is summarized in tab. \ref{times}: the above figures show how
stiff is the numerical problem of finding a dynamical solution of the NSE
in this framework. This is a physical situation in which several very
different time scales appear. In order to get an efficient and accurate code,
some approximations have to be made.
\begin{table}
\caption{\label{numbs}
Characteristic numbers implied in the dynamics of inertial modes of rotating
NS. Note that the definition of Chandrasekhar number is adapted to be of the order of $1$ for linear $l=m=2$ {\it r}-modes.}
\begin{tabular}{c|p{9.5 cm}|c}
\hline
\hline
 Numbers & Definition & Analytical expression\\
\hline
 Ekman & Ratio between $P$ and $T_v$ or between the viscous term and the
Coriolis force & $E\sim \nu/\l(n \omg R^{2}\r)$\\
 Rossby & Ratio between the typical velocities of the mode and of the
background fluid or between the non-linear term and the Coriolis force
 & $Ro \sim W/\l(R \omg\r)$\\
 Reynolds & Ratio between the Rossby and Ekman numbers or between the
non-linear and viscous terms & $Re \sim n W R/\nu$\\
 Chandrasekhar & Ratio between $T_v$ and $T_g$ or between the RR force and
the viscous term & $Ch \sim n^{2} G R^{9} \omg^{6}/\l(c^{7} \nu\r)$\\
\hline
\hline
\end{tabular}
\end{table}
\begin{table}
\caption{\label{times}
Typical values of the different time scales implied in the dynamics of
inertial modes of rotating NS.}
\begin{tabular}{c|p{6cm}|c|c}
\hline
\hline
 Time scale & Definition & Analytical expression & Typical value (or range)\\
\hline
 Acoustic & Travel of acoustic waves across the NS & $T_a\sim\frac{R}{c_s}$ &
$\sim 2.10^{-4}s$\\
 Inertial & NS's period (same order as the period of the linear inertial mode)
& $P=\frac{2 \pi}{\omg}$ & $> 2.10^{-3}s$\\
 Viscous  & Damping of inertial modes due to viscosity &
$T_v\sim 1/\l(2E\omg\r)$ & $> 10^7 P$\\
 Gravitational & Growing due to RR force &
$T_g\sim c^{7}P/\l(G M R^{4}\omg^{5}\r)$ & $\sim 10^8 P$\\
\hline
\hline
\end{tabular}
\end{table}

\section{Mass conservation, boundary conditions and approximations}
\label{approx}

To be consistent with the variables we chose in NSE, we have to write the mass
conservation equation with the dimensionless enthalpy. With the decomposition
explained at the end of the Section \ref{newtnl}, we have
\be
\tilde{H}\la\tau,\vec{\xi}\ra = \tilde{H}_0\la\vec{\xi}\ra + \alf\,\tilde{h}
\la\tau,\vec{\xi}\ra.
\ee

The first term $\tilde{H}_0$ is the background, the second $\tilde{h}$
corresponds to {\it the mode} itself and $\alf$ still quantifies the
non-linearity. This gives
\begin{eqnarray} \label{enthdf}
& \l(\pt_{\tau} + \pt_{\ph}\r)\,\tilde{h}+\l(\vec{V} \cdot \vnb \tilde{H}_0
+ \Gamma[n] \tilde{H}_0\,\vnb\cdot\vec{V}\r)+
\alf\,\l(\vec{V} \cdot \vnb
\tilde{h} + \Gamma[n] \tilde{h}\, \vnb\cdot\vec{V}\r) = 0
\end{eqnarray}
where $\Gamma[n] =\frac{{\d} \ln H}{{\d} \ln n} $. In the polytropic case,
$ P=k\,n^{\gam}$, $\Gamma$ is constant and reduces to $\Gamma=\gam-1$.
Exception done of the case $\gam\,=\,1$ where the enthalpy is a logarithm.
In the linear case, we have to neglect the last part of this
equation, {\it i.e.} to do $\alf = 0$ in Eq.(\ref{enthdf}). We shall now
describe some different choices that can be made to deal with this equation.
The reader can find in appendix \ref{implem} the algorithms to implement all
the following schemes in the framework of spectral methods.

\subsection{Solving the exact system of equation}

If one tried to solve numerically the exact Navier-Stokes and mass conservation
equations, one would have two main possibilities. The first would be to employ
an explicit scheme. But by this way, the Courant conditions for the following
of the acoustic waves would impose a time step very small compared with the
period of the star. This would almost forbid any hope to make evolutions during
durations long from the inertial modes point of view. The second way to proceed
would be to use an implicit scheme for the divergence free part (see the
appendix). Indeed, this would make it possible to take a time step not
too small compared with the period of the NS. But here the problem would be to
estimate the errors done. Hence, in a first study, some approximations can be
introduced to solve the problem in a easier way.

\subsection{The divergence free approximation}

To avoid the problem of solving acoustic waves for better studying inertial
modes, the easiest solution is to use the divergence free approximation. It
consists in replacing the usual mass conservation equation by $\vec{\nb} \cdot
\vec{V} = 0$. By this way the time step can be chosen larger than in the
solving of the exact system and then allows the following of the inertial modes
themselves. From a numerical point of view, this is a fast and robust
approximation quite useful in the exploring phase of the numerical work. Yet,
it should be noticed that this drastic approximation may be not too bad for
inertial modes in NS. Indeed, the $l=m=2$ \rmos which are the most interesting
for GW have a divergence free limit in the linear order. Furthermore, the
latest figures about bulk viscosity (Haensel {\it et al.} 2001\cite{haen01})
that are quite huge, could mean a fast damping of all modes, except of those
that are divergence free. Finally, note that in the linear limit of EE, this
approximation gives an evolution of the mode $\vec{V}$ that is independent of
the background star if it is rigidly rotating.

\subsection{The anelastic approximation}

Yet, instead of being quite ``savage'' with the equation and imposing the
divergence free condition, one could look for a cleverer way of doing
physics and think about inertial modes. A main feature of these modes is that
their frequency is of the same order as the angular velocity of the star in
which they occur. As their damping and growing times are larger than their
period ({\it cf.} Section \ref{charnumb}), one would like to take it for a
characteristic time scale. From practical and numerical points of view, it
means a time unit (or time step) choice not too small compared with the period,
in order not to waste memory and computational time calculating non relevant
physics. From a physical angle, the philosophy is to neglect acoustic waves by
assuming that time derivatives of the pressure and density perturbations do not
play a key role in the phenomenon. It can be done in a consistent way with the
anelastic approximation.\\

The anelastic approximation was first introduced in atmospheric physics by
Batchelor in 1953 \cite{bat53} and then derived from a rigorous scale analysis
by Ogura and Phillips in 1962 \cite{oguph62} who gave it its name. In
astrophysics, it appeared in 1976 in a paper by Latour {\it et al.}
\cite{lat76} concerning convection and was then widely used in that field and
others (such as stellar oscillations) where one can neglect temporal variations
of the perturbation in density but not necessarily spatial ones. For a recent
critical approach of this approximation within astrophysics, one can read
Dintrans and Rieutord 2001 \cite{dinrieu01} and Rieutord and Dintrans 2002
\cite{rieudin02}.\\

In the Eq.(\ref{enthdf}), anelastic approximation consists in neglecting
$\l(\pt_\tau + \pt_\ph \r)\tilde{h}$ which is the time derivative of the
enthalpy in the rotating frame. By this way, we cut acoustic waves and then
have
\begin{eqnarray} \label{anelenthdf}
& \vec{V} \cdot \vnb \tilde{H}_0 + \Gamma[n] \tilde{H}_0\,\vnb\cdot\vec{V}
+\alf\,\l(\vec{V} \cdot \vnb \tilde{h} + \Gamma[n] \tilde{h}\,
\vnb\cdot\vec{V}\r) = 0.
\end{eqnarray}

As a conclusion on this approximation, we would like to insist on a
particularity of the linear case. With anelastic approximation and
linearization of all equations, we obtain an equation for the mass
conservation that does not depend on the Eulerian perturbation of the enthalpy.
But, taking the curl of the linear EE (or NSE) equation gives an equation with
exactly the same feature (remember that the curl of a gradient is zero) with
an interesting additional fact: it neither depends on the Eulerian perturbation
of the gravitational potential. Furthermore, it is easy to verify that for a
background enthalpy depending only on $\xi$, the boundary condition is that
the radial velocity should vanish at the surface. Thus, the situation is that
finding the Eulerian perturbation of velocity should give exactly the same
result, whatever the hypothesis on the Eulerian perturbation of the
gravitational potential. It means, {\it we do not need Cowling approximation
with anelastic approximation in the Newtonian linear case}. Cowling
approximation would play a role only if we wanted to find what is the enthalpy
and what is the gravitational potential in the source term of the gradient
part of EE.

\subsection{Boundary conditions} \label{bcsec}

The only missing information is now the boundary conditions we chose. In
an actual NS, the fluid core is supposed to be surrounded by a more or less
rigid crust, an ocean and an atmosphere. Their physics is by itself quite a
complex subject (see for example Haensel 2001b \cite{haen01b}). But even for
a simple toy-model, for instance a crust made of only one type of
nuclei, depending on the temperature and on the amplitude of the motion of the
inner fluid, the physical state of the crust can be quite complex to describe,
something like icepack on the sea (see for instance
Lindblom {\it et al.} 2000 \cite{linowus00}). There is no need to explain how
difficult would be to translate this BC in a mathematical language... This is
the reason why, for the EE, we then chose to begin with
two different and extreme BC to get an idea of the limit cases. The first is
the free surface BC, {\it i.e.} the absence of any crust. The second is the
presence of a rigid crust at $\xi = 1$, if we take the inner radius of the
crust for the typical  length $R$. In the first case, one has
$\tilde{H}\mid_{\xi=1}=0$ and in the second $V_r\mid_{\xi=1}=0$. The numerical
ways to take into account those BC can be found in the appendix \ref{BC}. For
the NSE, as it was already said in Section \ref{charnumb}, in order to avoid
the need of more BC, we chose a {\it degenerate} shear viscosity of the
form $s=1-\xi^2$.

\section{Test and calibration of the code} \label{calib}

Now that we have presented our physical framework, we will focus on the code.
As it was already mentioned in the previous sections, it uses spherical
coordinates and spectral methods to solve NSE. More precisely, it solves
the equations coming from the decomposition of NSE into a potential and a
divergence free parts. We will not give more details here and send the reader
to the appendix for informations about the algorithms and spectral
methods. {\em The rest of this article, and the discussion about numerical
stability in the appendix, are devoted to the linear study. Non-linear work is
still in progress and will be described later}. Furthermore, in this section,
we only deal with rigid rotation in order to have some analytical solutions to
make tests with.

\subsection{Conservation of the energy}

The first test of the code was obviously the free evolution of the linear
$l=m$ \rmo in a rigidly rotating inviscid and incompressible fluid. The
vector defined in Eq.(\ref{rmodequa}) is indeed an eigenvector of EE whose
frequency in the inertial frame is $w_i = - \frac{(m-1)(m+2)}{m+1}$. Moreover,
note that the absence of radial component implies that both free surface and
spherical rigid crust BC are automatically satisfied. Yet, for all the
preliminary calculations using the divergence free approximation, the code
was built to work with the rigid crust BC.\\

\begin{figure}
\begin{center}
\includegraphics[width=6.9cm,height=8.6cm,angle=-90]{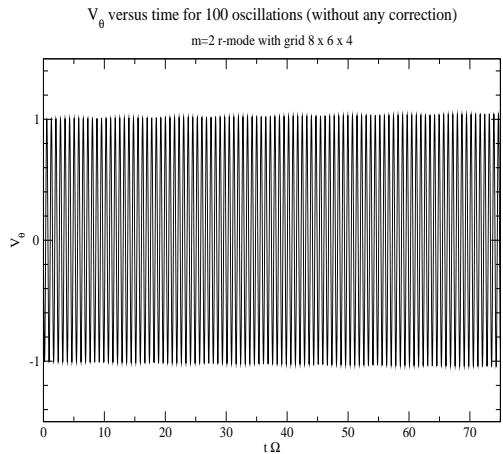}
\caption[]{\label{vth-fig}
Time evolution of the $\th$ component of velocity on the equator for the linear
$m=2$ \rmo in a incompressible and rigidly rotating fluid. The run is supposed
to last exactly 100 times the period of this mode which is $\frac{3}{4 \omg}$.
The beating phenomenon which can be seen is due to the resolution of the
graphical tool. See the power spectrum in figure (\ref{pw-fig}).}
\end{center}
\end{figure}

\begin{figure}
\begin{center}
\includegraphics[width=6.9cm,height=8.6cm,angle=-90]{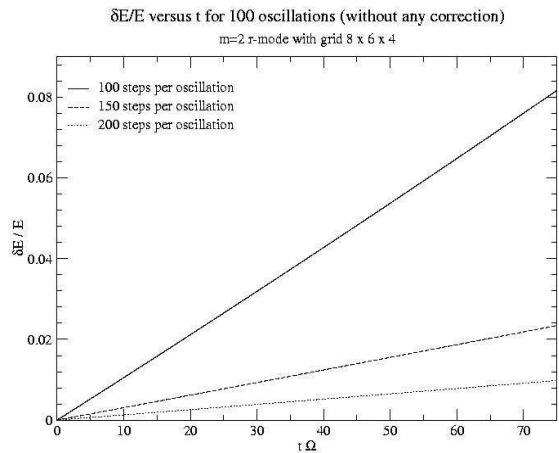}
\caption[]{\label{en100-fig}
Time evolution of the relative error in the energy before any improvement of
the conservation of energy. The different straight lines correspond to runs
with the same duration (100 periods of the $m=2$ linear {\it r}-mode), same
spatial grid ($8 \times 6 \times 4$ points) but different time steps. We see
the linear variation of these errors with time. It shows that exception done
of round-off errors (which are so small that they do not appear here), the
main error is deterministic and then well controlled.}
\end{center}
\end{figure}

\begin{figure}
\begin{center}
\includegraphics[width=6.9cm,height=8.6cm,angle=-90]{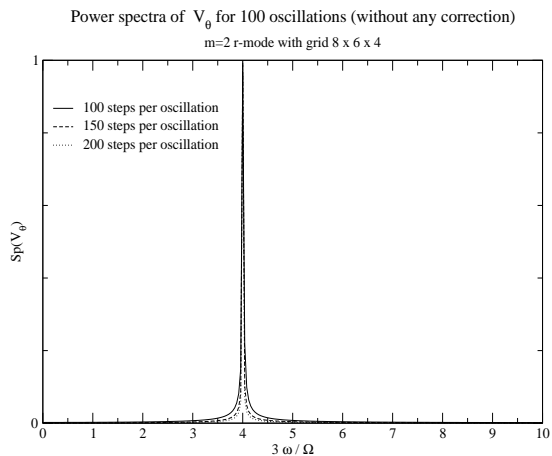}
\caption[]{\label{pw-fig}
Power spectra of component drawn in figure (\ref{vth-fig}) and of the same
component calculated with 150 or 200 steps per oscillation. See also figure
(\ref{en100-fig}).}
\end{center}
\end{figure}

The great advantage of spectral methods is to change any linear spatial
operation in linear algebra calculations, which can be done
exactly. As the velocity of Eq.(\ref{rmodequa}) has a very small
number of coefficients in the reciprocal space, it is easy to verify (for 
example by directly looking at the time evolution of those coefficients) that
there are only errors due to round-off and to time discretization.
In figure (\ref{vth-fig}) is illustrated the time evolution of $V_\th$ at the
equator for the $m=2$ linear {\it r}-mode. The spatial lattice was of the
shape $(8 \times 6 \times 4)$ for
$(N_r,N_\th,N_\ph)$\footnote{In fact, symmetries of the spherical harmonics
are used and the effective value of $l_{max}$ is roughly twice the number of
points in $\th$.} with 100 time steps per period of the mode. The duration of
this run was chosen for the evolution to last exactly 100 times the expected
period. This calculation was done on a DEC Alpha Station with 500 MHz processor
and took 217 seconds of CPU time. Comparison of the amplitude at the first step
and at the last steps shows the growth of the amplitude due to errors that
come from the time discretization. This is easier to see in
figure (\ref{en100-fig}) where the time evolution of the error in energy
appears. The different curves illustrate results obtained with the same grid
and duration, but in runs with different numbers of time steps per oscillation.
Here are pictured results for 100, 150 and 200 [see also associated power
spectra in figure (\ref{pw-fig})]. The last two calculations on the same
computer took respectively 269 and 342 seconds of CPU time.\\

\begin{figure}
\begin{center}
\includegraphics[width=6.9cm,height=8.6cm,angle=-90]{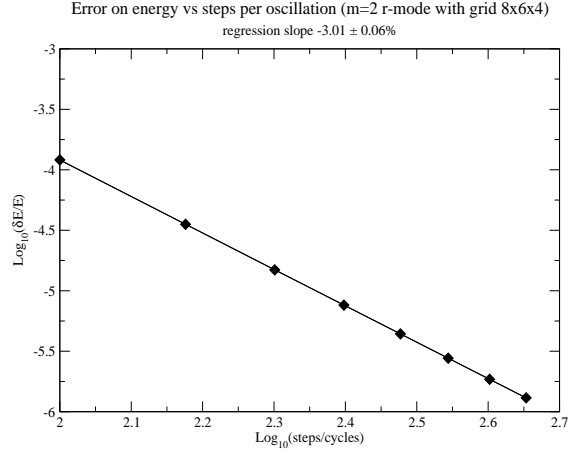}
\caption[]{
\label{erren-fig} On this picture clearly appears the fact the error exactly
comes from the adopted numerical scheme with an error that vanishes as the
third power of the time step. The relative error and the number of steps per
oscillation are both in logarithmic scales and the regression slope is then
$-3.01\,\pm\,0.06\%$.}
\end{center}
\end{figure}

In figure (\ref{erren-fig}), we drew for several runs the relative error in the
energy per oscillation of the mode, versus the number of time steps per
oscillation. Both are in logarithmic scales. This error appears to exactly
varies as $\Dt t^3$ and then as the inverse of the cube of the number of steps
per period (regression slope of $-3.01$). This is due to the second order
scheme we use to solve the NSE or EE. Indeed, when the energy is calculated,
the error of order $\Dt t^2$ done on the velocity is multiplied by a source
term linear in $\Dt t$.\\

As we are now only dealing with the linear code, we will not discuss the
stability of the non-linear code, nor give test pictures corresponding to
modes with azimuthal numbers different from $2$. Indeed, there is a
coupling between different values of $m$ only in the non-linear versions
of NSE and EE. The modes that are the most interesting for GW are $m=2$
modes and in the following, we will always choose this kind of initial
data. Nevertheless, the same calibration tests for the other linear
$l=m$ \rmos were done and gave the same power law relations between
$\frac{\dt E}{E}$ and the number of steps per oscillation. Finally, we also
verified that the error in the phase of the modes is proportional to the
square of the number of steps per period. The relative errors in the phase for
the $m=2$ mode with 100, 150 and 200 steps per oscillation were
respectively $1.64\,10^{-3},\,7.33\,10^{-4}\,\text{and}\,4.12\,10^{-4}$.
It is easy to verify that the logarithms of these numbers are on a straight
line with a slope very close of $2$.\\

We shall see now how we tried to improve the conservation of energy. For the
free {\it r}-mode, the energy is expected to be exactly conserved. Furthermore,
we know our numerical error in the velocity is of the order of $\Dt t^2$.
Remember that it comes from the second order scheme used to calculate
source terms:
\be \label{extrasource}
S^{j+1/2}= \frac{3 S^j\,-\,S^{j-1}}{2}.
\ee

\begin{figure}
\begin{center}
\includegraphics[width=6.9cm,height=8.6cm,angle=-90]{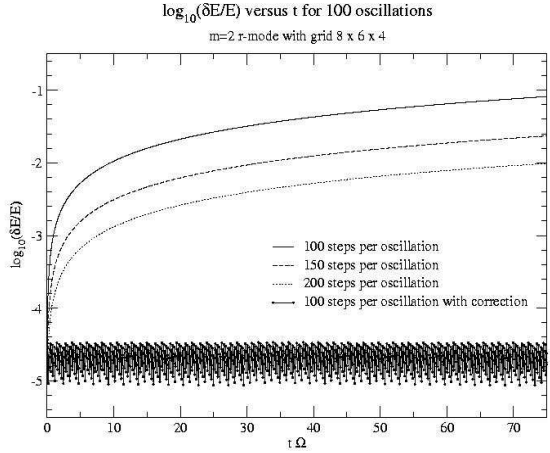}
\caption[]{\label{en100-cor-fig}
Here are pictured the same curves as in fig.(\ref{en100-fig}), but in
logarithmic scales. We added the curve corresponding to a run with 100 steps
per period of the $m=2$ linear \rmo with the correction of the conservation
of energy described toward Eq.(\ref{extrasourcemod}).}
\end{center}
\end{figure}

The basic idea is thus to modify this quantity, in the coefficients space, with
some additional term of the order of $\Dt t^2$, in such a way to retrieve the
conservation of the energy without changing the error in the velocity. We then
modified the Eq.(\ref{extrasource}) and took
\be \label{extrasourcemod}
S^{j+1/2}= \frac{ (3+\eps) S^j - (1+\eps) S^{j-1}}{2}
\ee
where $\eps$ is of the order of $\Dt t^2$. It is calculated using values of
the energy at the last two instants and in such a way to ``impose'' on the
energy to be conserved. Obviously, when there are forces such as viscosity or
RR, their power has to be taken into account in the energy balance equation. In
fig.(\ref{en100-cor-fig}) are the same curves as in fig.(\ref{en100-fig})
but in logarithmic scales. We added the result of this ``improved conservation
of energy'' for the run with 100 steps per oscillation. As the correction is
local in time (done at almost each time step), it leads to a time
independent error that should be able to remain the same as long as one could
wish. This calculation took 320 seconds of CPU time.

\subsection{Modes driven to instability in the rigidly rotating case}

\begin{figure}
\begin{center}
\includegraphics[width=6.9cm,height=8.6cm,angle=-90]{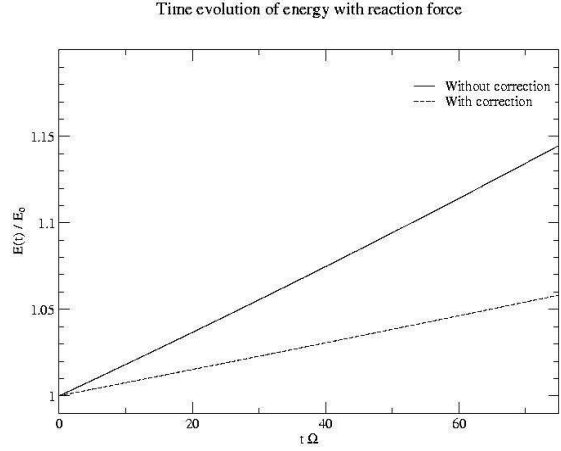}
\caption[]{
\label{rr-fig}
Time evolution of the ratio between energy and initial energy in an inviscid
and incompressible rigidly rotating fluid with a RR force corresponding to what
should exist in a NS with $M= 1.4 M_{\odot}$, $R = 10$ km and
$\omg\simeq(2\pi)\,500 Hz$. The first curve corresponds to the basic code and
the second one to the code with improved conservation of energy. The expected
final value is about $1.057$. The difference between this analytical
calculation and the corrected numerical result is less than $10^{-3}$.}
\end{center}
\end{figure}

After trying to improve the conservation of energy for the free case, we did
tests with a RR force included. For instance, we switched on the RR force in
order to drive toward instability the linear $m=2$ {\it r}-mode. As it has
already been implied in Section \ref{charnumb}, we did it by using a modified
version of the formulae (\ref{radreac}) (see this section). For these
calculations, we compared numerical results with analytical calculations, both
reached with the same approximation. We took for the background star a
homogeneous NS with $M = 1.4 M_{\odot}$, $R = 10$ km and $\omg \sim
\l(2 \pi\r) 500 Hz$. In fig.(\ref{rr-fig}) appear two time evolutions of the
ratio between the energy and the inertial energy for 100 oscillations of the
mode. The first calculation was done without the enhanced conservation of
energy and the second with this improvement. The analytical calculation gives
for the ratio a final value of the order of $1.0575$ whose difference with the
improved numerical value, $1.0582$, is less than $10^{-3}$.\\

\begin{figure}
\begin{center}
\includegraphics[width=6.9cm,height=8.6cm,angle=-90]{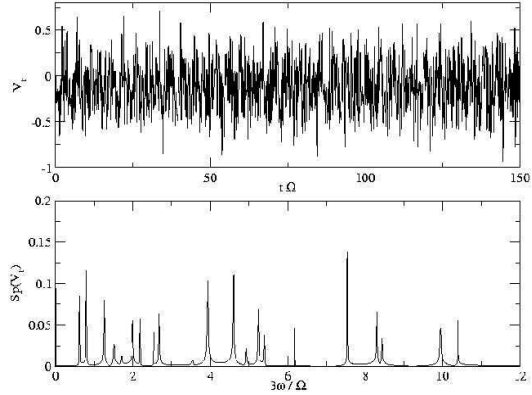}
\caption[]{
\label{rnoisevr}
Time evolution of the radial component of velocity in an inviscid and
incompressible rigidly rotating fluid with Gaussian noise for initial data and
associated power spectrum. A huge RR force acts, but as expected, no polar
counter part of the mode grows.}
\end{center}
\end{figure}

\begin{figure}
\begin{center}
\includegraphics[width=6.9cm,height=8.6cm,angle=-90]{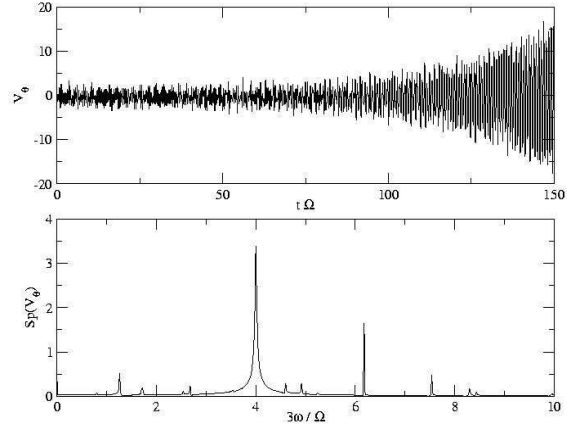}
\caption[]{
\label{rnoisevt}
Time evolution of the $\th$ component of velocity on the equator in an
inviscid and incompressible rigidly rotating fluid and associated power
spectrum with same initial data in fig.(\ref{rnoisevr}). Here, the $l=m=2$
linear \rmo is clearly driven to instability by the RR force. As expected, its
frequency is $w=\frac{4}{3}\omg$. Others peaks are just marks of the initial
noise.}
\end{center}
\end{figure}

\begin{figure}
\begin{center}
\includegraphics[width=6.9cm,height=8.6cm,angle=-90]{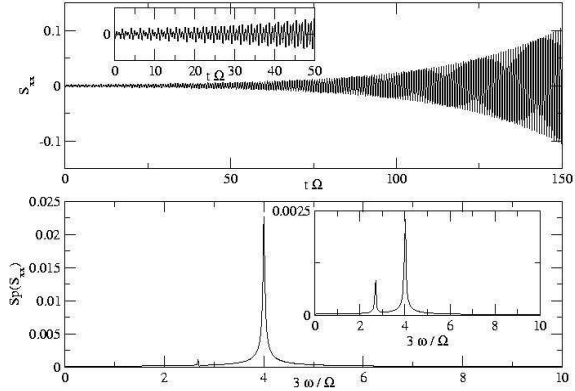}
\caption[]{
\label{noisij}
Time evolution of one of the two independent components of the $S_{ij}[t]$
tensor that appears in the RR force. This calculation was done during the
same run as the results in fig.(\ref{rnoisevr}) and (\ref{rnoisevt}). We
can also see the almost monochromatic associated spectrum. Nevertheless, there
is another frequency in this tensor that is not driven to instability. It can
be seen more easily in the zoom of the time evolution or in the corresponding
power spectrum.}
\end{center}
\end{figure}

The next step was to switch on the RR force, but without the linear \rmo
for initial data, even if we still assumed that the fluid was divergence free.
Instead, we chose for the initial velocity a random Gaussian noise with a first
moment equal to $0$ and a second moment equal to $1$. It should be noticed
that in this case and others where the initial energy is not only distributed
within quite low frequency modes, the above method to improve the conservation
of energy is no longer appropriate and we shall not use it. Moreover, even if
this method does not change the angular momentum when applied to the $m=2$
part of the velocity, it can if one naively applied it to the $m=0$ part. But,
as we already mentioned, here we only deal with the linear code and mainly with
$m=2$ velocities. And in this case, we exactly control the error done, without
changing the angular momentum. We drew in figures
(\ref{rnoisevr}) and (\ref{rnoisevt}) the time evolutions of the $r$ and $\th$
components of the velocity and their power spectra. In fig.(\ref{noisij}) is
illustrated the time evolution of one of the two independent components of the
$S_{ij}$ tensor, with the associated power spectrum. These calculations were
done with a lattice of the shape $(12 \times 8 \times 4)$, with $200$ steps per
oscillation and lasted $200$ times the period of the $m=2$ {\it r}-mode. The
included RR force corresponds to what should exist in a NS with
$M = 1.4 M_{\odot}$, $R = 10$ km and $\omg \simeq (2 \pi)\,1050 Hz$. Thus,
this calculation is not really physical (due to the use of the slow rotation
approximation) but just aims at giving a qualitative idea of what should happen
with physical values. As expected, there is an axial mode (no radial velocity)
driven to instability with exactly the frequency ($w=\frac{4}{3}\omg$) and
coefficients ($l=2$) of the {\it r}-mode. Moreover, a single look at the figure
(\ref{noisij}) shows that even when the velocity is mainly noisy, the tensor
that plays the key-role in the instability is quite smooth. In this
figure, we also drew a zoom corresponding to $t \omg < 50$ and the associated
power spectrum. It shows two frequencies in this component of the tensor. One
of them is the unstable \rmo with $3 w/\omg =4$ and the other (with
$3 w/\omg \sim 2.70$) {\it disappears} with a longer run, as the scale is
adapted to the growing mode. Yet, even in the spectrum of the full evolution,
a trace of it can still be seen. We achieved exactly the same features with
others {\it noisy} initial conditions, for instance a {\it Dirac kick} into the
NS\footnote{We mean an initial velocity equal to $0$ anywhere except in an
arbitrary point.}. We also did the same calculations with other spatial
lattices (up to $64 \times 48 \times 4$), and this result did not change at
all.

\subsection{Test of the anelastic approximation}

The last calculations we did in a rigid and Newtonian background were to test
the effect of the anelastic approximation. The idea was to compare the previous
results obtained with the divergence free approximation and a rigid crust BC
with results coming from the same initial conditions but with the anelastic
approximation and the free surface BC. For the rest of this section,
the background star is a polytrope with $\gam=2$.\\

\begin{figure}
\begin{center}
\includegraphics[width=6.9cm,height=8.6cm,angle=-90]{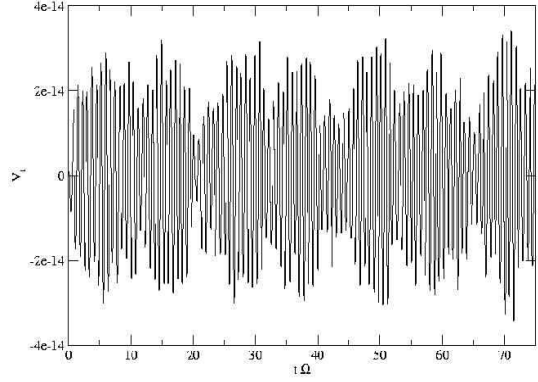}
\caption[]{
\label{vr-an-fig}
Time evolution of the radial component of the velocity in the equatorial plane
in $\xi=0.5$ with the linear $m=2$ \rmo for initial data. The background star
is a rigidly rotating $\gam=2$ polytrope and the calculation is done with the
anelastic approximation and without any RR force. The run lasts 100 times the
period of the {\it r}-mode. There should be no radial velocity ({\it cf.} text)
and we see it comes only from round-off errors (double precision
calculation with initial values of the order of the unity).}
\end{center}
\end{figure}

First, we took for initial data the linear $l=m=2$ {\it r}-mode and let it
freely evolve. Here again, the BC do not play any role. The spatial lattice was
$(8 \times 6 \times 4)$ with 100 time steps per period of the mode. The time
evolution of $V_\th$ at the equator [same as in fig.(\ref{vth-fig})] did not
show any difference with the divergence free case, even for the power spectrum.
But this is quite obvious to understand when looking at both the
Eq.(\ref{anelenthdf}) and the EE. Indeed, we see that the linear {\it r}-mode,
which is divergence free and has no radial component, is still an eigenvector
of this system of equations. Then taking it for initial data in the divergence
free case or in the anelastic approximation should give exactly the same
evolution. We verified [see fig.(\ref{vr-an-fig})] that the radial component of
the velocity does not grow and neither does its divergence. This calculation
took 344.6 of CPU time without any optimization and with a very basic solving
of the anelastic equation.\\

The next step was the driving toward instability of a mode with noise for
initial data. We took exactly the same conditions (duration, lattice, initial
data) as in the divergence free case, exception done of the boundary
conditions. For these calculations and all that follows with the anelastic
approximation, we will use the free surface condition that is automatically
satisfied (see the appendix for more details). The figures (\ref{tfvran-fig}),
(\ref{tfvthan-fig}) and (\ref{tfsijan-fig}) that show respectively the time
evolutions of $V_r$, of $V_{\th}$ and of a component of $S_{ij}$ give a kind 
of summary of the results:
\bi
\item [-] the radial velocity still remains noisy and does not grow;
\item [-] the $\th$ component grows in the same way as in the divergence free
case. There is a difference between the final amplitudes that comes from the
fact that the coupling to the RR force depends on an integral on the whole
star of a function that is proportional to the density (and then depends on the
EOS). Note that the power spectra have the same secondary peaks as in the
divergence free case.
\item [-] the spectrum of the $S_{ij}$ tensor shows that there is only one
unstable mode which has again the frequency of the linear {\it r}-mode.
\ei

The conclusion is then that the anelastic approximation gives results very
close of those obtained in the divergence free case. This is due to the fact
that, in spite of the presence of all modes in the initial conditions, the
only growing mode is the \rmo that has no radial velocity and is divergence
free.

\begin{figure}
\begin{center}
\includegraphics[width=6.9cm,height=8.6cm,angle=-90]{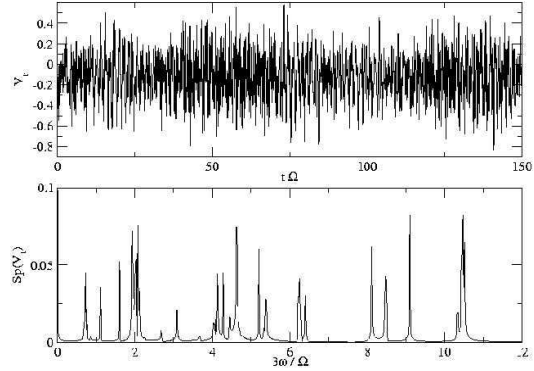}
\caption[]{
\label{tfvran-fig}
Time evolution of the radial component of velocity with Gaussian noise for
initial data and its power spectrum. The background is a polytrope with
$\gam=2$ and the anelastic approximation is used. As in fig.(\ref{rnoisevr})
a huge RR force acts, but as the star is rigidly rotating no polar counter
part of the mode grows.}
\end{center}
\end{figure}
\begin{figure}
\begin{center}
\includegraphics[width=6.9cm,height=8.6cm,angle=-90]{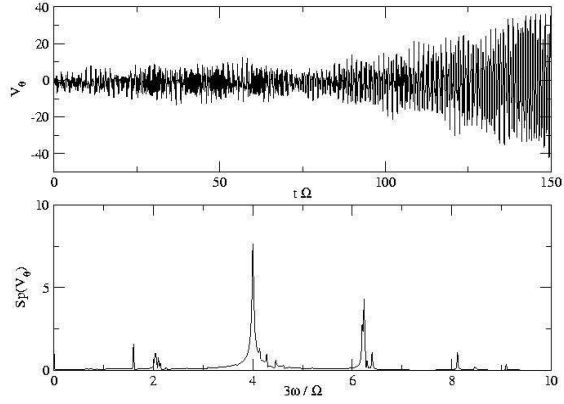}
\caption[]{
\label{tfvthan-fig}
Time evolution of the $\th$ component of velocity and associated power spectrum
for the same calculation as in fig. (\ref{tfvran-fig}). The only difference
with the divergence free case [{\it cf.} fig.(\ref{rnoisevt})] is the final
amplitude. Yet, this is not due to the anelastic approximation but to the
effect of the EOS on the coupling coefficient of the RR force.}
\end{center}
\end{figure}
\begin{figure}
\begin{center}
\includegraphics[width=6.9cm,height=8.6cm,angle=-90]{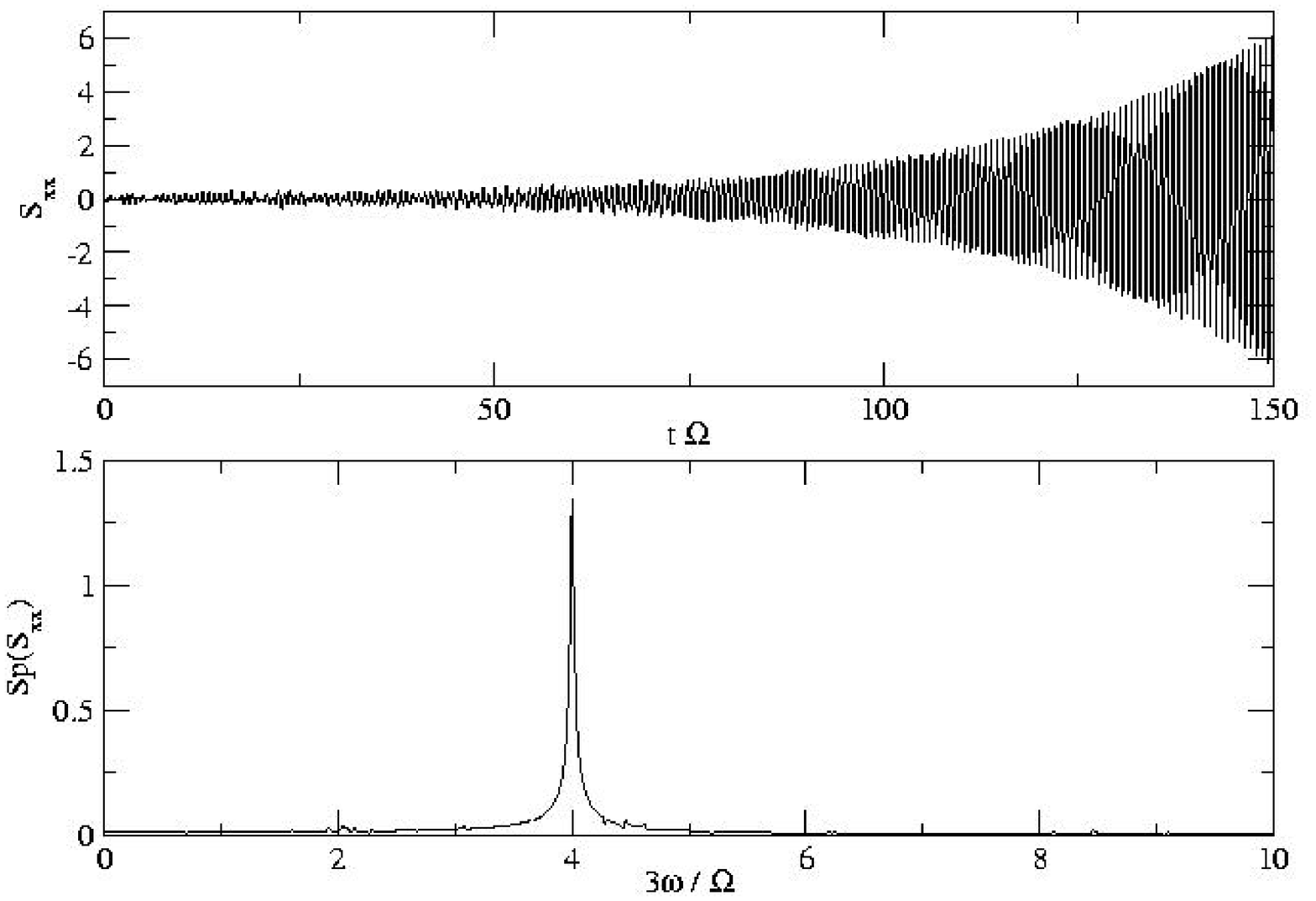}
\caption[]{
\label{tfsijan-fig}
Time evolution of one of the two independent components of the $S_{ij}[t]$
corresponding to the same run as the results in fig.(\ref{tfvran-fig}) and
(\ref{tfvthan-fig}). Here can really be seen the fact that there is only one
unstable mode with the frequency of the linear {\it r}-mode and that
the anelastic approximation does not change this feature.}
\end{center}
\end{figure}

\section{Differential rotation} \label{difrot}

As we already mentioned it in the introduction, Kojima first noticed (see
Kojima 1998 \cite{koj98}) that relativistic \rmos may be quite different from
what they are in a Newtonian background due to frame dragging. Yet, since the
main reason for this difference is the modification of the NSE and the
possible appearance of what is called a {\it continuous} spectrum (see
Ruoff {\it et al.} 2001 \cite{ruoff01}, \cite{ruoff01b} and Beyer {\it et al.}
1999 \cite{beyer99}), the same may append even in the Newtonian case if the
star is not rigidly but differentially rotating. And there are several reasons
why a NS may not be in rigid rotation. First, the birth conditions of
the NS themselves. Secondly, the non-linear coupling of modes. Then, a
possible drift induced by the existence of a magnetic field. As we are here
only dealing with linear hydrodynamics and tests of the code, we will not give
more details about those processes (and send the reader to the
following articles for more details: Spruit 1999 \cite{spr99},
Rezzolla {\it et al.} 2000 \cite{rezls00}, 2001 \cite{rezlms01} and
\cite{rezlms01b}, and Schenk {\it et al.} 2002 \cite{sch02}). What we have
done is just to take as given that the background star is differentially
rotating with quite an arbitrary law and to look at the influence of this law
on the existence of the modes. Nevertheless, instead of asking the question
``Is there any \rmo left in a differentially rotating NS?'' that is not well
defined, we decided to try to answer to two different and more precise
questions:
\bi
\item [-] is there anything growing when a RR force is applied on noise in a
differentially rotating background?
\item [-] what does happen to the linear \rmos if they are chosen to be the
initial data in such kind of background?
\ei

Some lights on these questions are in the following subsections, but we shall
begin with some words about the modifications implied on the equations by
differential rotation.

\subsection{Modifications}

Assuming differential rotation of the background star involves slight
modifications of what we said in the Section \ref{dim}. First, terms
coming from the spatial derivatives of the non constant
$\vec{\omg}[\xi,\th,\ph]$ must be added in the NSE. Secondly, we shall
now specify what exactly means $\omg$ in the definition of dimensionless
variables [{\it cf.} Eq.(\ref{var})].\\

Concerning $\omg$, we chose to take for a time unit the inverse of its value
at the equator. This is uniquely determinate due to the fact we have always
assumed that the rotation law is of the form $\omg[r,\th,\ph] =
\omg[r \sin[\th]]$ or of the form $\omg[r,\th,\ph] = \omg[r]$. The first case
corresponds to what must be this law for the background to be stable with
respect to the Newtonian EE, and the second case is in a way {\it inspired}
by GR even if it is not a solution of the full Newtonian EE. For more details
see Section \ref{grcas}.\\

In the dimensionless NSE, the modifications induced by differential rotation
are quite simple. First, $\pt_{\ph}$ is replaced with $\tilde{\omg}[r,\th]
\pt_{\ph}$ where $\tilde{\omg}[r,\th]$ is the dimensionless profile
of rotation. Then, we have to add new terms coming from $\vec{\rm e}_\ph
\,r \sin[\th]\,\vec{V}\cdot \vnb \l(\tilde{\omg}[r,\th]\r)$ that are in
the spherical orthonormal basis
\be \label{drotterm}
\left|\begin{array}{ll}
&0\\
&0\\
V_r r \sin[\th] \pt_r{\tilde{\omg}}
& +\,\, V_{\th} \sin [\th] \pt_{\th}{\tilde{\omg}}.\\
\end{array}
\right.
\ee

\subsection{Noise with huge RR force}

Once again, our goal was to stay as close as possible of the basic and well
understood situation to minimize the number of unknowns. Thus, we took noisy
initial data, put it in a differentially rotating background, switched on
the RR force and looked to what was to happen. By this way, the question was
not to look for the existence of {\it r}-modes, but simply to look for the
existence of modes driven to instability by the RR force in a non rigidly
rotating background.\\

The only difference with the basic study done in the case of rigid rotation was
that we had to choose a law for $\omg$. The first choice was very simple and
corresponded to a background stable with respect to the Newtonian EE. It was
of the form
\be \label{nwdifrot}
\tilde{\omg}[r,\th] = {\cal W} \l(1 + \beta_n\,r^2 {\sin[\th]}^2 \r)
\ee
where $\beta_n$ was a constant depending on the run and ${\cal W}$ calculated
to have $\tilde{\omg} \la 1,\frac{\pi}{2} \ra = 1$. As explained above, we also
tried a law {\it inspired} by GR:
\be
\tilde{\omg}[r,\th] = {\cal W} \l(1 + \beta_{gr}\,r^2 \r)
\ee

or laws coming from the already quoted article by
Karino {\it et al.} \cite{karino01}:
\be \label{kldifrot}
\tilde{\omg}[r,\th] = \frac{{\cal W}\,{\beta_L}^2}
{ r^2 {\sin[\th]}^2 + {\beta_L}^2}
\ee

or
\be
\tilde{\omg}[r,\th] = \frac{{\cal W}\,{\beta_v}}
{{\l(r^2 {\sin[\th]}^2 + {\beta_v}^2 \r)}^{1/2}}.
\ee

Since in all these laws, none of the free parameters $\beta$ corresponds to a
physical variable, we will not give quantitative results. It will be done in
the relativistic study (and then in another article) where there is a parameter
that physically quantifies the way the equations are far from the Newtonian
EE: the compactness of the star. Here we will only discuss in a qualitative
way the results obtained in all the previous cases and give a very
representative example: what happens in a $\gam=2$ polytrope with anelastic
approximation (free surface) and the rotation law given by
the Eq.(\ref{nwdifrot}) with $\beta_n=0.4$. Exception
done of these choices, the calculation was done with exactly the same
conditions as in the last study with noisy initial data in the previous
section.\\

\begin{figure}
\begin{center}
\includegraphics[width=6.9cm,height=8.6cm,angle=-90]{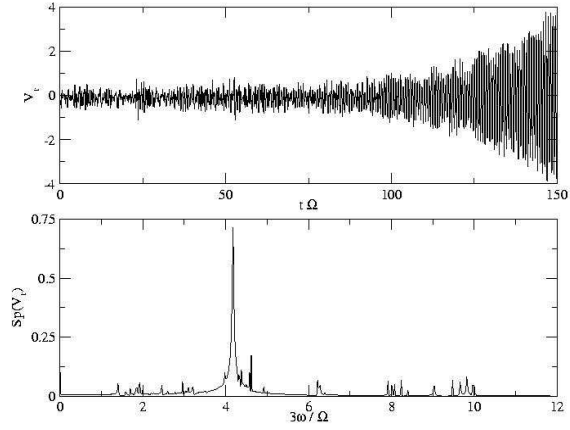}
\caption[]{
\label{rnoisevr-ane}
Time evolution of the radial component of velocity in a $\gamma=2$ polytrope
with anelastic approximation and Gaussian noise for initial data. A huge RR
force acts (something like what should exist in a star with angular
velocity equal to $1000$ Hz) and the background star is assumed to be
differentially rotating with the law corresponding to $\beta_n=0.4$. We see
that a polar counter part of the mode now grows.}
\end{center}
\end{figure}

\begin{figure}
\begin{center}
\includegraphics[width=6.9cm,height=8.6cm,angle=-90]{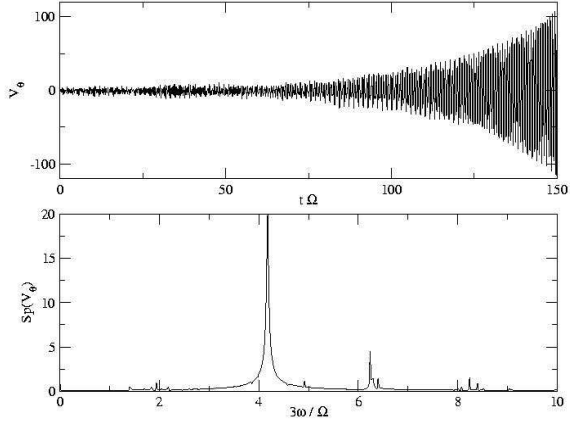}
\caption[]{
\label{rnoisevt-ane}
Time evolution of the $\th$ component of velocity in the same calculation as
in fig.(\ref{rnoisevr-ane}). The associated power spectrum shows that there
is one mode driven to instability that is the same as in
fig.(\ref{rnoisevr-ane}) and has a frequency very close of the expected
frequency of the {\it r}-mode. See also the fig.(\ref{rnoisesij-ane}).}
\end{center}
\end{figure}

\begin{figure}
\begin{center}
\includegraphics[width=6.9cm,height=8.6cm,angle=-90]{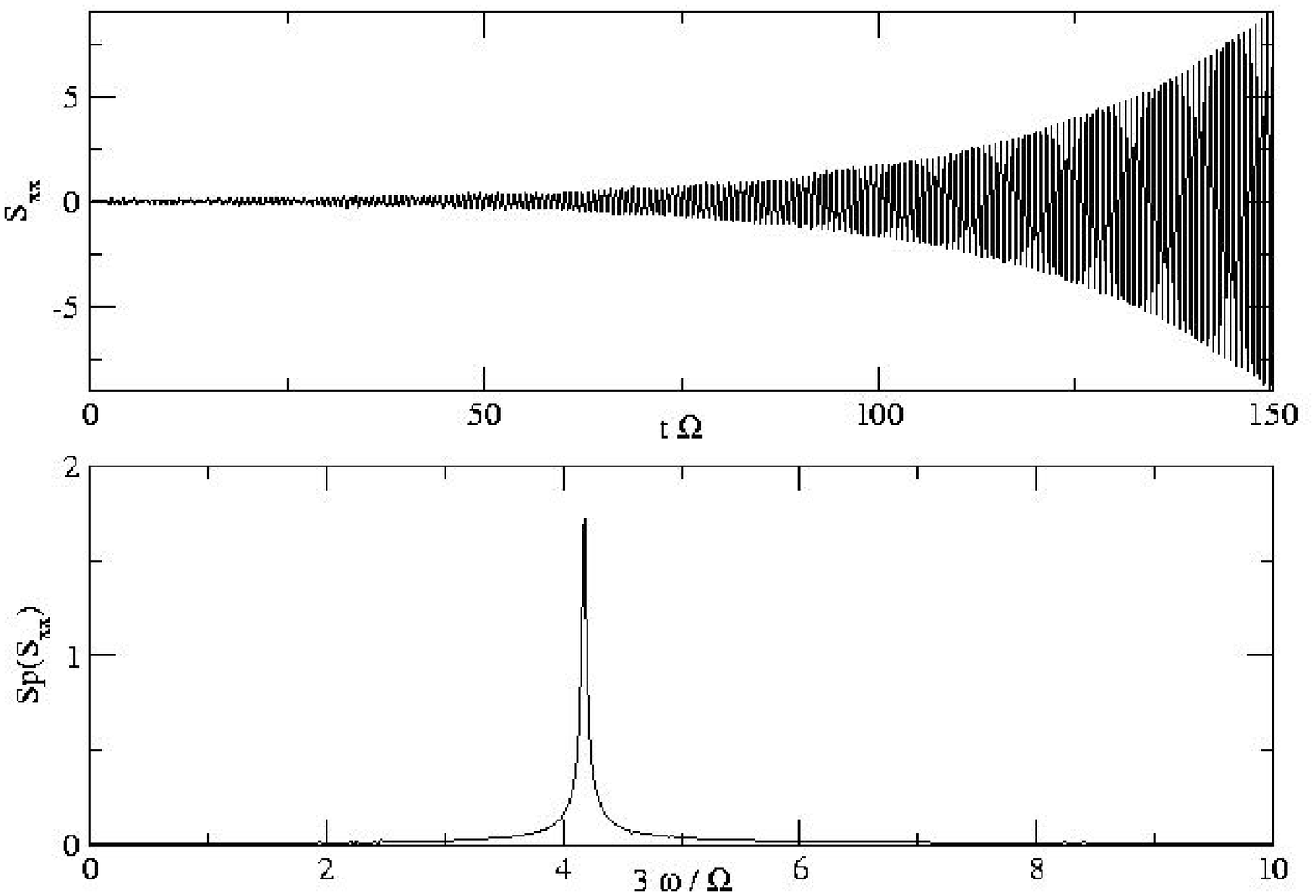}
\caption[]{
\label{rnoisesij-ane}
Time evolution of one of the two independent components of the $S_{ij}[t]$
tensor that appears in the RR force. This calculation was done during the
same run as the results in fig.(\ref{rnoisevr-ane}) and (\ref{rnoisevt-ane}).
We can see the almost monochromatic associated spectrum with the same
frequency as in the previous spectra of the unstable mode.}
\end{center}
\end{figure}

The main difference with the case of rigid rotation is that the mode
that grows is no longer purely axial. Indeed, we can see on
fig.(\ref{rnoisevr-ane}) that the radial velocity is now also driven to
instability. Moreover, comparing power spectra in this figure and in
fig.(\ref{rnoisevt-ane}) and (\ref{rnoisesij-ane}) shows that this is really
a single mode with frequency very close of the frequency of the {\it r}-mode.
By {\it a single mode}, we mean that this is exactly the same frequency for
several physical quantities and several positions in the star. This is not
a sufficient condition to talk about ``discrete spectrum'', but it is a
necessary condition. This unavoidable existence of a polar part of the velocity
in a differentially rotating Newtonian star with barotropic EOS should be
compared with the results achieved by Lockitch {\it et al.} \cite{lockandf01}
in the relativistic framework. Indeed, in GR, the main reason for the coupling
between axial and polar parts of the velocity is the frame dragging that is
imitated in a Newtonian framework by differential rotation. Finally, note that
the value of the frequency in the Newtonian case is depending on the way we
chose to normalize $\omg$. Then, to summarize all our calculation, we can say
that we observed that {\it the polar counter part of the mode appears as soon
as there is differential rotation, and whatever the chosen law}. Yet, the more
the law for $\omg$ is far from the rigid case, or in more pragmatic way the
greater the free parameter is, the more the unstable mode has a polar counter
part.

\subsection{Free evolution}

The other question we asked was: ``What does happen to linear \rmos if they are
put into a differentially rotating background?''. The idea in taking such kind
of initial data was to have something quite close of an eigenvector, assuming
if there is one it should be quite similar to the linear {\it r}-mode.\\

\begin{figure}
\begin{center}
\includegraphics[width=6.9cm,height=8.6cm,angle=-90]{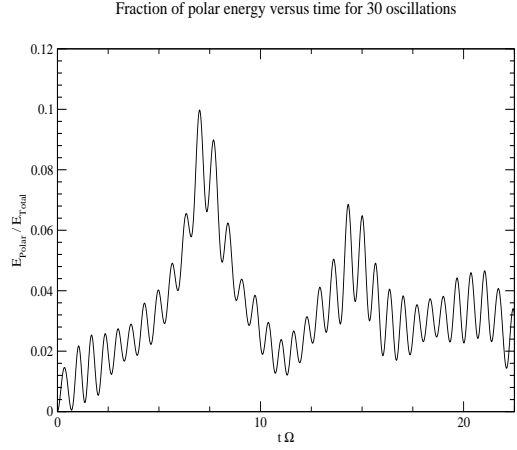}
\caption[]{
\label{freeenpol}
Time evolution of the ratio between energy contained in the polar part
of the velocity and the total energy of the mode in the anelastic approximation
with a free surface. The background star is assumed to be a $\gam=2$ polytrope
differentially rotating with the law corresponding to $\beta_n=0.4$. A
viscous term coming from the degenerate viscosity with $E_s=5.\,10^{-5}$ is
included. At the beginning of the evolution, there is energy coming from the
purely axial initial data (the linear {\it r}-mode) and going to the polar
part. Then this energy oscillates from the polar part of the velocity to the
axial part and back with a constant mean value.}
\end{center}
\end{figure}

\begin{figure}
\begin{center}
\includegraphics[width=6.9cm,height=8.6cm,angle=-90]{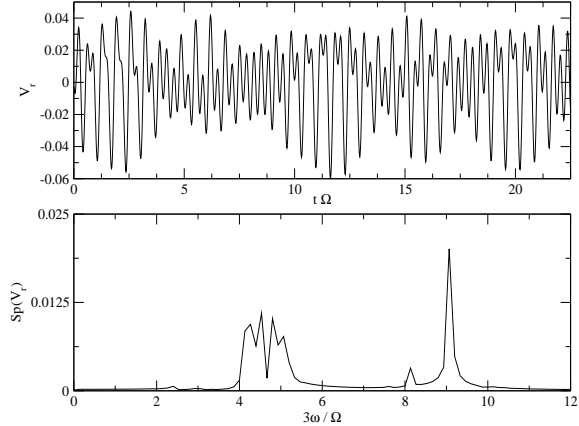}
\caption[]{
\label{freevrdiv}
Time evolution of the radial component of velocity with the anelastic
approximation and a free surface. This calculation was done
during the same run as in fig.(\ref{freeenpol}). The initial data are the
linear \rmo that is no longer an eigenvector and radial velocity quickly
appears. Several different modes are present.}
\end{center}
\end{figure}

Once again, we did not make quantitative calculations and postponed it
to the GR case. As in the case of noisy initial data driven toward
instability by a RR force, the free evolution of the linear \rmo
showed the growth of a polar part of the velocity. It can be seen on
figures (\ref{freeenpol}) and (\ref{freevrdiv}) that respectively show
the time evolution of the ratio between the polar energy and the total
energy and the time evolution of the radial velocity at the point of
coordinates $\l(\xi=\frac{1}{2},\th=\frac{\pi}{2},\ph=0\r)$. This
results comes from a calculation done with a spatial lattice of the
shape $(64,48,4)$, the anelastic approximation (with a free surface)
and a rotation law given by the Eq.(\ref{nwdifrot}) with
$\beta_n=0.4$. For reasons that will be explained later, we included
degenerate viscosity ({\it cf.}  Section \ref{bcsec}) with an Ekman
number $E_s\,=\,5.\,10^{-5}$ in order to regularize the solution. The
evolution was done to last $30$ periods of the linear {\it r}-mode
with $100$ steps per oscillation. In the figure (\ref{freeenpol}), we
see that, after a while, the ratio between the energy in the polar
part of the mode and the total energy reaches a kind of stationary
state with a coupling between different modes. The existence of this
``hybrid final state'' was verified during other runs with other
physical conditions and is once again to compare with results achieved
in GR by Lockitch {\it et al.} \cite{lockandf01}.\\

\begin{figure}
\begin{center}
\includegraphics[width=6.9cm,height=8.6cm,angle=-90]{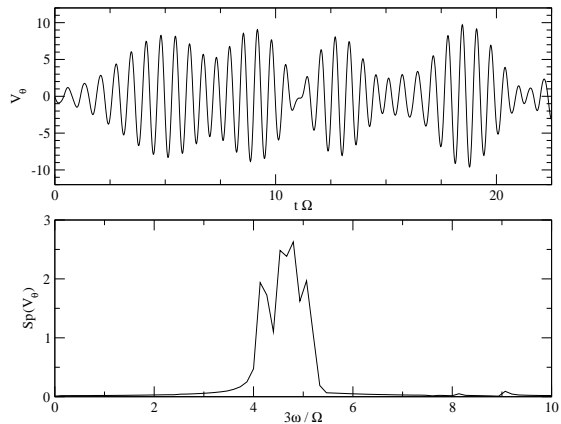}
\caption[]{
\label{freevtdiv}
Time evolution of the $\th$ component of velocity in the same
calculation as in fig.(\ref{freevrdiv}). The curve and the associated power
spectrum show that there are several modes, but only one of them appears
in the $S_{ij}$ tensor with quite a huge amplitude. Moreover, it has a
frequency very close of the frequency of the linear {\it r}-mode. See also
the fig.(\ref{freesijdiv}). Note finally that the same frequency also appears
in the spectrum of the radial velocity.}
\end{center}
\end{figure}

\begin{figure}
\begin{center}
\includegraphics[width=6.9cm,height=8.6cm,angle=-90]{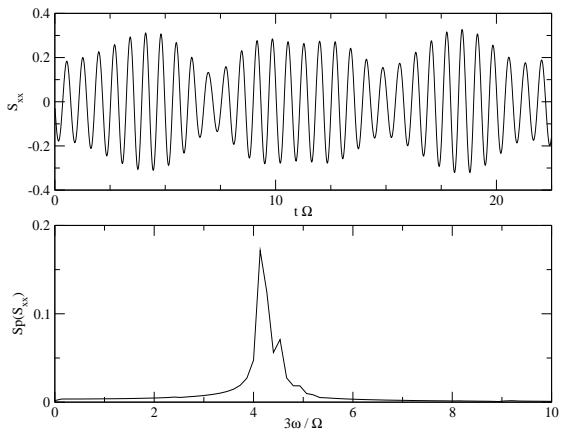}
\caption[]{
\label{freesijdiv}
Time evolution of one of the two independent components of the
$S_{ij}[t]$ tensor that appears in the RR force. This calculation was done
during the same run as the results in fig.(\ref{freevrdiv})
and (\ref{freevtdiv}). We can see there seems to be one main frequency and a
second one of smaller importance. The first one has exactly the same frequency
as the unstable mode that appears in the previous figures and calculations
where the RR force was on.}
\end{center}
\end{figure}

Concerning the existence of modes, looking simultaneously at
fig.(\ref{freevrdiv}), (\ref{freevtdiv}) and (\ref{freesijdiv}) shows
that apparently one single mode mainly appears in the reaction force (or
in the component of the corresponding tensor) even if both axial and polar
part of the velocity contain several modes. Yet, the spectrum of the
$S_{ij}$ tensor is quite noisy due to the fact that the RR force is not
switched on and that the run is short. We verified this feature during other
calculations. The conclusion is that for small values of the $\beta$
parameter (these values are depending of the chosen rotation law), the main
effect of differential rotation on a ``free linear {\it r}-mode'' is to give
it a polar counter part and to widen its spectrum. But for larger
values of the parameter, even if the velocity's spectrum becomes very noisy,
the $S_{ij}$ tensor is always less noisy. Yet, here we will not give more
details about this points, the linear Newtonian study being not really
interesting from the NS point of view.\\

\begin{figure}
\begin{center}
\includegraphics[width=6.9cm,height=8.6cm,angle=-90]{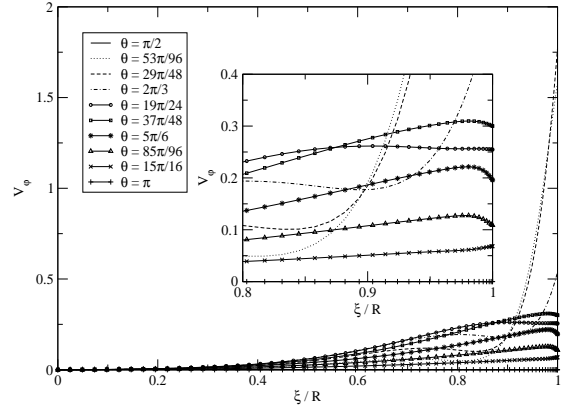}
\caption[]{
\label{shapdr}
$\ph$ component of the velocity versus the radius for several
values of $\th$. We can see the main part of the velocity is concentrate
in a region near the surface and close of the equator. This ``snapshot''
of the profile was taken at a moment when the derivative versus the radial
coordinate of the velocity is quite huge at the surface.}
\end{center}
\end{figure}

\begin{figure}

\begin{center}
\includegraphics[width=6.9cm,height=8.6cm,angle=-90]{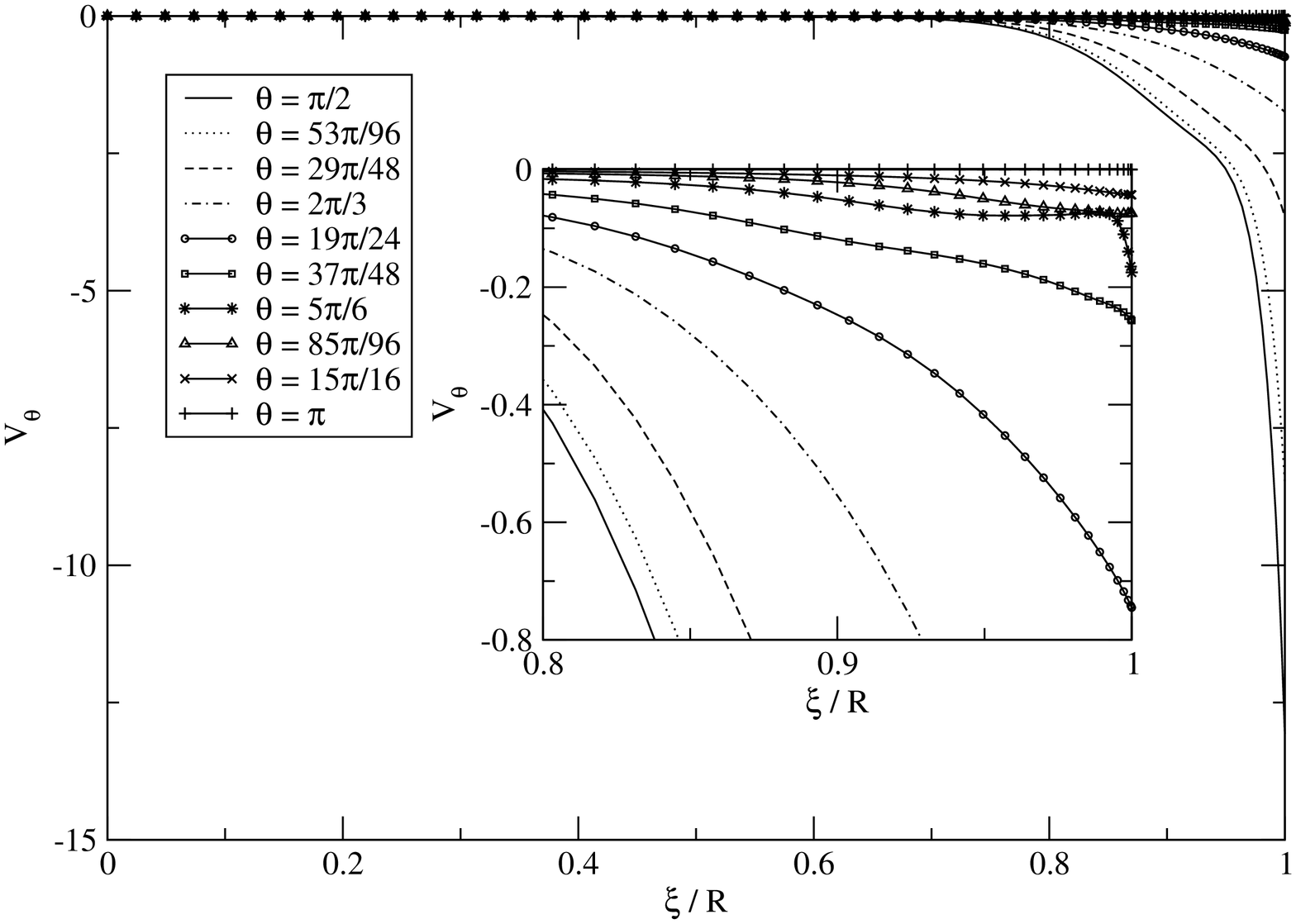}
\caption[]{
\label{shapdr2}
$\th$ component of the velocity versus the radius for several
values of $\th$. Here the concentration of the motion is higher than
in the previous calculation. $\beta_n$ was $0.4$ in fig.(\ref{shapdr})
and is now $0.8$. Moreover, this calculation lasted three time longer.
Nevertheless, note that here we did not choose to draw the profile
when the derivative is huge and just drew it at the final instant.}
\end{center}
\end{figure}

To end with this section, we will mention a quite amazing feature found in all
our calculations and illustrate it with an example coming from the previous
free run of a linear \rmo put in the differentially rotating background. We
always found that quite fast (in something like $15$ periods of the linear
{\it r}-mode) the main part of the velocity is most of the time concentrated
in a region close of the surface of the star and of the equator. This is
illustrated in the figure (\ref{shapdr}) where we drew the shape of the $\ph$
component of the velocity versus the radius of the star for several values of
the $\th$ angle. Note that we got the
same results with different BC and even when we add viscosity to regularize
the solution (and this is the case in this calculation) and to be sure this is
not a numerical artifact. In fig.(\ref{shapdr2}) appears the shape of $\th$
component for an evolution done with the same grid, viscosity and initial data
but with $\beta_n=0.8$. With such a huge value of the parameter that governs
the rotation law, there is a lot of different modes that appear in the velocity
spectrum. But what seems to be the most important result is the very
high concentration of the velocity near the surface. This is analogous to the
results achieved by Karino {\it et al.} in 2001 \cite{karino01}. In the
conclusion, we will shortly discuss what is, in our point of view, a possible
important repercussion of this result.\\


\section{Strong Cowling approximation in GR} \label{grcas}

NS are among the most relativistic macroscopic objects made of usual matter
in the Universe. Moreover, the instability of \rmo is due to GR. It is then
quite natural to try to study \rmos in the GR framework. The problem is that
it is quite difficult to evolve dynamically the full GR and hydrodynamics
equations taking into account GW. This is the reason why we decided to use the
strong Cowling approximation at the beginning.\\

Here we will only give some preliminary results obtained in the GR case. A more
complete and physical study will be in a following article in preparation. Our
goal is just to show that the above conclusions concerning the appearance of a
polar part of the velocity and the ``concentration of motion'' for Newtonian
differentially rotating NS still hold in the framework of relativistic rigidly
rotating NS.\\

As we are studying slowly rotating NS, which remain spherical, a very
convenient and good approximation is to assume that the $3$-space is
conformally flat (see Cook {\it et al.} 1996 \cite{cook96}). Furthermore, as
we are using the strong Cowling approximation, there is no GW even
without this approximation. Then, in unit $c=1$, the infinitesimal
space-time interval can be written
\begin{eqnarray}
& {ds}^2 = - \l({N[r]}^2-{a[r]}^2\,r^2\,{\sin[\th]}^2\,{N^{\ph}[r]}^2\r) {dt}^2
- 2 {a[r]}^2\,r^2\,{\sin[\th]}^2 \,N^{\ph}[r] dt\,d\ph
+ {a[r]}^2\,{d\vec{r}}^2 \nonumber
\end{eqnarray}
where, following the $3+1$ formulation, $N[r]$ is the lapse function
and $N^{\ph}[r]$ the third contravariant component on the spherical
coordinate basis of the shift vector, ${a[r]}$ is the conformal factor
and finally ${d\vec{r}}^2$ is the infinitesimal interval in the
$3$-space. In this equation, we use the isotropic gauge and the slow
rotation limit to assume that all these {\it unknown} functions are
only depending on $r$ (and this is the reason why, in the Newtonian
study, we called ``inspired by GR'' the law of differential rotation
in which the functions are only depending on $r$). In Hartle 1967
\cite{hartl67}, the system of coordinates is not the same as the one
we use, but it is easy to show that the conclusions are identical since the
mapping between the two systems is quite trivial. Furthermore, from
the same article we know that ${N^{\ph}[r]}$ is of the first order in
$\omg$.\\

To get the equations of motion, the relativistic equivalent of EE, we write
the energy tensor of a perfect fluid
\be
T^{\mu\nu}=(\rho+p)U^{\mu}U^{\nu}+p g^{\mu\nu}
\ee
where $\rho$ is the energy density, $p$ the pressure and $U^{\mu}$ the
$4$-velocity of the perfect fluid. Then, we apply the conservation of energy:
\be
\nb_{\mu}T^{\mu\nu}=0.
\ee

If we define the generalized enthalpy:
\be
{\d} H \widehat{=} \frac{{\d} P}{\rho+p},
\ee

it gives
\be \label{REE}
U^{\mu}\,U^{\nu}\,\nb_{\nu} H\,+\,\nb^{\mu} H\,+\,U^{\nu}\,\nb_{\nu} U^{\mu}\,
=\,0.
\ee

It is well-known that due to thermodynamics, the system of equations obtained
by adding the baryonic number conservation $\nb_{\mu}\l(n\,U^{\mu}\r)$ (where
$n$ is the baryonic density) to those four equations is a degenerate system.
Then, we chose to work with the baryonic number conservation and the
projections on the $3$-space of Eq.(\ref{REE}).\\
 
Following the Newtonian case, we just have to add Eulerian perturbations to
the rigid rotation and to linearize the equations with respect to both the
amplitude of these perturbation and $\omg$. For the generalized enthalpy, the
definition of the perturbation is obvious, but for the full $4$-velocity, we
use the well-known results about rigid rotation of relativistic stars and
write it as
\be
U^{\mu}[t,r,\th,\phi] \, = \frac{1}{N[r]}
\left| \begin{array}{ll}
1+\dt U^0[t,r,\th,\phi] & \\
{\l(\frac{N[r]}{a[r]}\r)}^2\,\dt U^r[t,r,\th,\phi] & \\
{\l(\frac{N[r]}{a[r]}\r)}^2\,\frac{\dt U^{\th}[t,r,\th,\phi]}{r} & \\
\omg+ {\l(\frac{N[r]}{a[r]}\r)}^2\,\frac{\dt U^{\ph}[t,r,\th,\phi]}
{r \sin[\th]} & \\
\end{array} \right.
\ee

Note that in this equation, $\dt U^0$ is not a dynamical variable but is
determined according to the constraint that $U^{\mu}$ is a $4$-velocity
($U^{\mu}U_{\mu}=-1$). Furthermore, $\dt U^r, \dt U^{\th}$ and $\dt U^{\ph}$
are not contravariant components of a $4$-vector but convenient variables that
are the components of a $3$-vector on the orthonormal basis associated with the
spherical system of coordinates for the flat $3$-space. It enables us to write
the motion equations in a way very similar to the Newtonian EE. Indeed, writing
the $3$-velocity on the orthonormal basis associated with the spherical
coordinates as
\be
\vec{V}\widehat{=}\,\left|\begin{array}{l}
\dt U^r\\
\dt U^{\th}\\
\dt U^{\ph}\\
\end{array}
\right.
\ee
and defining on the same basis
\be
\oa{\varpi}[t,r,\th,\phi]\widehat{=} 
\left\{ \begin{array}{ll}
\l(\omg\,-\,N^{\ph}[r]\r)\cos[\th] & \\
-\,\sin[\th]\,\l(\omg\,-\,N^{\ph}[r]\,+\,A[r]\,\r) & \\
0 & \\
\end{array}
\right.
\ee
with $A[r]\widehat{=}\,\l(\omg\,-\,N^{\ph}[r]\r)\l(\frac{{\d \ln }
[\frac{a[r]}{N[r]}]}{{\d \ln}[r]}\r)\,-\,\frac{r\,N^{\ph'}[r]}{2}$
where $'$ is the derivative versus the radial coordinate, we have
\be
\l(\pt_t + \omg\, \pt_{\ph}\r)\,\vec{V}+\,2\,\vec{\varpi}\wedge\vec{V}+
\vnb h=0.
\ee

This equation is very close of the Newtonian EE, the main difference being that
the $3$-vector that appears instead of $\vec{\omg}$ is now depending on the
coordinates (as in the case of differential rotation) but no longer parallel
to the rotation axis.\\

Concerning the baryonic number conservation, writing it in a Newtonian like
way, we have in the slow rotation limit
\be
\l(\pt_t + \omg\, \pt_{\ph}\r)\,\tilde{n}  + {\l(\frac{N[r]}{a[r]}\r)}^2 \dv
\l(\tilde{n}\,\vec{V}\r) = 0
\ee
where $\tilde{n}$ is defined as $\tilde{n} \widehat{=} n\,N^2\,a$, $n$ being
the baryonic number density. The natural generalization of the anelastic
approximation is then
\be
\dv \l(\tilde{n}\,\vec{V}\r) = 0.
\ee

As we are using the strong Cowling approximation, the background star
appears in the equations of motion only as ``external'' (from the point
of view of the mode) data. For any relativistic calculation, what we do is
to calculate the background configuration using the already existing code
illustrated in Bonazzola {\it et al.} 1993 \cite{bona93} and then to use the
resulting lapse, shift, conformal factor and their derivatives with respect
to the radial coordinate in our equations.\\

\begin{figure}
\begin{center}
\includegraphics[width=6.9cm,height=8.6cm,angle=-90]{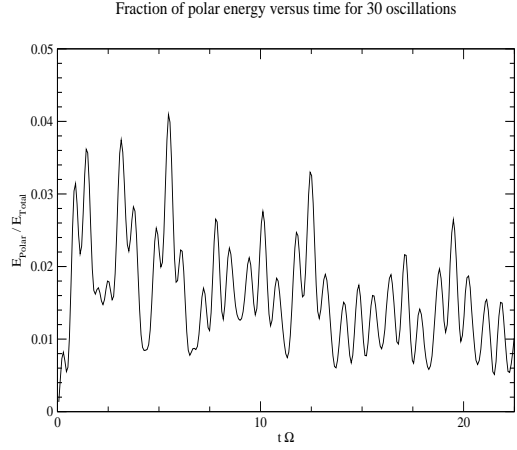}
\caption[]{
\label{rgfenpol}
Time evolution of the ratio between energy contained in the polar part
of the velocity and the total energy of the mode with the strong Cowling and
anelastic approximations and the free surface BC. The background star is a
$\gam=2$ relativistic rigidly rotating polytrope with $1.74$ solar masses and a
radius equal to $12.37$ km. The ratio between its kinetic energy and its mass
energy was $10^{-16}$. We see that in spite the fact the initial data are the
linear Newtonian {\it r}-mode that is purely axial, as soon as the calculation
begins, a polar counter part of the mode appears as implied by
Lockitch {\it et al.} \cite{lockandf01}.}
\end{center}
\end{figure}

\begin{figure}
\begin{center}
\includegraphics[width=6.9cm,height=8.6cm,angle=-90]{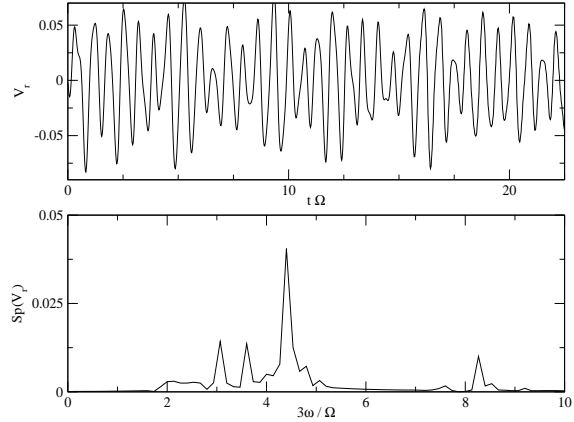}
\caption[]{
\label{rgfvrdiv}
Time evolution of the radial component of velocity at the point
$(\frac{1}{2},\frac{\th}{2},0)$ for the same calculation as in
fig.(\ref{rgfenpol}). The initial data are the linear \rmo that is no longer an
eigenvector and radial velocity appears very fast. The power spectrum of the
time evolution shows several peaks, but one of them has the same frequency
as the modes of fig.(\ref{rgfvtdiv}) and (\ref{rgfsijdiv}) and is the
relativistic analog of the {\it r}-mode. But it is no longer purely axial.}
\end{center}
\end{figure}

\begin{figure}
\begin{center}
\includegraphics[width=6.9cm,height=8.6cm,angle=-90]{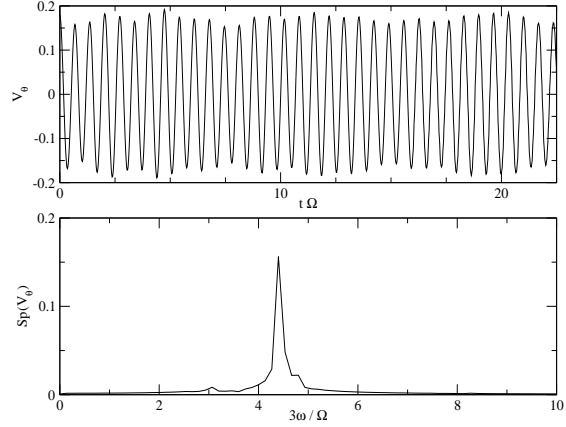}
\caption[]{
\label{rgfvtdiv}
Time evolution of the $\th$ component of velocity in the same
calculation as in fig.(\ref{rgfvrdiv}). The curve and the associated power
spectrum show that there is one mode that also appears in the polar part
with a frequency very close of the frequency of the linear {\it r}-mode. See
also the fig.(\ref{rgfsijdiv}).}
\end{center}
\end{figure}

\begin{figure}
\begin{center}
\includegraphics[width=6.9cm,height=8.6cm,angle=-90]{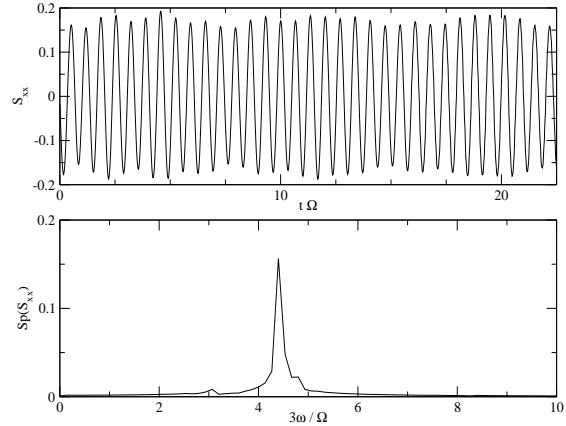}
\caption[]{
\label{rgfsijdiv}
Time evolution of one of the two independent components of the
$S_{ij}[t]$ tensor that appears in the RR force. This calculation was done
during the same run as the results in fig.(\ref{rgfvrdiv})
and (\ref{rgfvtdiv}). We can see the almost monochromatic associated
spectrum with exactly the same frequency as the mode that appears in the
previous figures.}
\end{center}
\end{figure}

\begin{figure}
\begin{center}
\includegraphics[width=6.9cm,height=8.6cm,angle=-90]{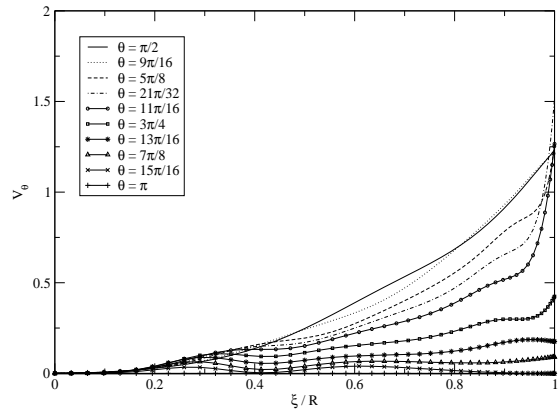}
\caption[]{
\label{shaprgdr}
$\th$ component of the velocity versus the radius for several values of $\th$.
We can see the kind of concentration of the motion near the surface and the
equatorial plane. This calculation was done with the strong Cowling and
anelastic approximations and the free surface BC. The background star is a
$\gam=2$ relativistic rigidly rotating polytrope with $1.74$ solar masses and
a radius equal to $12.37$ km. The initial data are the Newtonian
linear $l=m=2$ {\it r}-mode. This figure corresponds to the velocity after a
time equal to $15$ oscillations of the linear $l=m=2$ {\it r}-mode.}
\end{center}
\end{figure}

Here we will only give one single example, showing we have in this relativistic
case with the strong Cowling approximation results similar to those obtained
in the case of the Newtonian differential rotation. We took a $\gam=2$
polytrope with $1.74$ solar masses and a radius equal to $12.37$ km. The star
was very slowly rotating (the ratio between its kinetic energy and its mass
energy was about $10^{-16}$) and we took for a time unit the inverse of the
angular velocity. The anelastic approximation with the free surface BC was used
and to regularize the solution, we added a degenerate viscosity
with $E_s=5.\,10^{-5}$ once the final equations were written in a Newtonian
way. We insist on the fact that this viscous term does not come from
relativistic calculations and just aims at regularizing the solution. In
figures (\ref{rgfenpol}), (\ref{rgfvrdiv}), (\ref{rgfvtdiv}),
(\ref{rgfsijdiv}) and (\ref{shaprgdr}) are plotted exactly the same quantities
as in the Section \ref{difrot}. The conclusions for this preliminary
relativistic calculation are the same as in the Newtonian case: a polar
counter part of the velocity appears from the beginning of the evolution and
most of the time, the velocity is concentrate near the equator and the
surface. 

\section{Conclusion} \label{conc}

This article deals with a new hydrodynamical code based on spectral
methods in spherical coordinates. This first version of the code was
written to study inertial modes in slowly rotating NS, but it can
easily be modified to include fast rotation. As it was explained in
the introduction, the slow rotation limit is a very good approximation
for a first step. The present version of the code can be used to study
both the Newtonian case and the relativistic case with the so-called
strong Cowling approximation. Here, we have given the main algorithms
it uses and shown some tests of the linear version. The way we
overcame the numerical instability is also explained.\\

Furthermore, in order to work with very different time scales, we have
introduced an approximation to deal with mass (or baryonic number)
conservation that proved itself quite robust and useful to study
inertial modes: the anelastic approximation. Even if this
approximation is not a necessary one and can be abandoned in further
studies, it is very useful to have an idea of the main properties of
the inertial modes in slowly rotating stars. The results presented in
this article were obtained in the linear case using very basic models
of NS, such as the divergence free case with a rigid crust
($V_r\mid_{r=1}=0$) or a $\gam=2$ polytrope with anelastic
approximation and free surface. Deeper studies in the GR framework
with more quantitative results on the effects of EOS, of
stratification of the star and of BC will follow. Despite the naive
aspect of our studies, some common features appeared. Indeed, whatever
the approximation (divergence free or anelastic) with different BC
(rigid crust for the divergence free case and free surface for the
anelastic approximation), we saw the appearance of a polar part in the
mode, as soon as the background is Newtonian and differentially
rotating, or relativistic and rigidly rotating (in the latter case,
this phenomenon is due to the frame dragging). Whether in the case of
noisy initial data with a radiation reaction force acting on them, or
in the case of the free evolution of an initial {\it linear r}-mode,
from the very beginning the energy of the polar part is at least one
per cent of the energy of the axial part.  This result is compatible
with the analytical work by Lockitch {\it et al.} \cite{lockandf01}
who proved, in the relativistic framework, the existence of a
non-vanishing polar part of the velocity in a sequence of NS with a
barotropic equation of state and decreasing rotation
rate. Furthermore, another interesting result that we obtained is the
``concentration of the motion'' near the surface that quickly appears:
after less than $15$ periods of the linear \rmo in the calculations
done up to now. Adding viscosity has shown that this is a robust and
physical feature. Moreover, it seems to neither depend on the physical
conditions or on the BC. This phenomenon is very similar to the
results found by Karino {\it et al.} in 2001 \cite{karino01} with a
eigenvectors calculation.  But, if this is really a singularity of the
derivative of the velocity near the surface, as it seems to be in our
time evolutions, it would mean that a mode calculation with finite
difference schemes can only give quantitative results completely
depending on the number of points. Indeed, with this scheme, there is
an intrinsic numerical viscosity that depends on the resolution.
Moreover, if this phenomenon is verified in following studies in GR,
the stratification of the NS or the physics of the crust could be very
important to know if inertial modes are relevant for GW emission.

\acknowledgments

We would like to thank E. Gourgoulhon and B. Carter for carefully
reading this paper. The numerous remarks of the anonymous referee were
also very useful for us in trying to improve this article.
Then, L. Blanchet and J. Ruoff gave us the opportunity to have
friendly and fruitful discussions. Finally, we would like to thank the
computer department of the observatory for the technical assistance.\\

\appendix

The numerical algorithm adopted to solve the NSE (or the EE) is based on
the pseudo-spectral methods (PSM) widely used in hydro or MHD problems. Before
explaining this algorithm in more details, we will begin with a short summary
of the PSM in order to make more evident the peculiarity of the solving
of vectorial equations like NSE. Then, follows a second appendix that deals
explicitly with EE and aims at explaining the way we implemented the
approximation done on the mass conservation equation.

\section{Spirit of the pseudo-spectral methods} \label{spirit}

Our group developed algorithms and routines library (see Bonazzola {\it et al.}
1990, 1999 \cite{bonam90},\cite{bonagm99})\footnote{In 1980, one of us (S.B.)
started to build a library of routines based on spectral methods to solve
PDE in different geometries. Today, this library contains more than 700
routines written in FORTRAN 70 and 90 languages. These routines are highly
hierarchised and allow us to assemble codes in modular way. We call this
library ``Spectra''. A part of this library (the highest in the hierarchy)
was written in $C^{++}$ language by J.A. Marck and E. Gourgoulhon in order
to allow the use of an object oriented language. This library is called
``Lorene''. The code described in this section uses the ``Spectra'' library
and is written in FORTRAN 90 language.} allowing us to solve partial
differential equations (PDE) in different geometries, mainly in domains
diffeomorphic to a sphere. First, with an example of scalar PDE, we will look
at the singularities contained in operators expressed in spherical like
coordinates [see for instance the different components of the NSE:
Eq.(\ref{NSEDC})] and at our choice of spectral basis. Then, we shall
discuss the further difficulties that arise in vectorial PDE and explain
the way we overcome them.

\subsection{The scalar heat equation}

Consider the heat equation:
\be \label{heat}
\frac{\pt F}{\pt t}=\alpha \Delta F+S
\ee
where $\Delta F=\frac{\pt^2 F}{\pt r^2}+\frac{2}{r}\frac{\pt F}{\pt r}
+ \frac{1}{r^2} (\frac{\pt^2 F}{\pt \th^2} +\frac{\cos \th}{\sin \th}\,
\frac{\pt F}{\pt \th} +\frac{1}{\sin \th^2} \frac{\pt^2 F}{\pt \ph^2})$
is the Laplacian in spherical coordinates, $\alpha$ the heat
conductivity supposed to be constant, and $S$ a source term
(that may include non-linear terms if they exist). The idea of spectral
methods is to look for the solution of the Eq.(\ref{heat}) on the form
\be \label{PSM}
F[r,\th,\ph,t] = \sum_{n,l,m=0}^{\infty} F_{nlm}[t]\,H^l_n[r]\,
K^m_{l}[\th]\,L_m[\ph]
\ee
with ($0\leq r \leq 1;0 \leq\th \leq \pi;0 \leq \ph \le 2 \pi$)
and where $H^l_n[r], K^m_l[\th], L_m[\ph]$ are a well chosen complete
set of functions.  The problem is then to find the time evolution of
the coefficients $F_{n l m}[t]$.\\

In a spherical geometry, it is quite natural to choose $L_m[\ph]
=\exp^{im\ph}$ and $K^m_l[\th]=P^m_l[\th]$, where $P^m_l[\th]$ are
the Legendre functions. With these choices, the Eq.(\ref{heat})
can be written
\begin{eqnarray} \label{heatleg}
& \frac{\pt F_{lm}[r,t]}{\pt t}-\alpha\l(\frac{\pt^2 F_{lm}[r,t]}{\pt r^2}
+\frac{2}{r}\frac{\pt F_{lm}[r,t]}{\pt r }-\frac{l(l+1)}{r^2}
F_{lm}[r,t]\r)=S_{lm}[r,t]
\end{eqnarray}
where $S_{lm}[r,t]$ are the Fourier-Legendre coefficients of the function
$S$ at the radius $r$ and the instant $t$.\\

In order to handle the singularity at $r=0$, we shall consider separately the
cases $l=0$ , $l=1$ and $l > 1$:
\begin{itemize}
\item [-] For $l=0$, we use the fact that a $C^1$ function symmetric with
respect to the inversion $r \rightarrow -r$ has its first derivative that
vanishes at least as $r$ at $r=0$. Therefore, for such a function, the term
$\frac{\pt^2 F}{\pt r^2} + \frac{2}{r} \frac{\pt F}{\pt r}$ is regular. Even
Chebyshev polynomials $T_{2n}[r]$ have this property, and the choice
$H^0_{n}[r] = T_{2n}[r]$ ($n \ge 0$) then satisfies the regularity
conditions.
\item [-] For $l=1$, it is almost the same, but the final choice is
$H^1_n[r]=T_{2n+1}[r]$.
\item [-] The case $l >1$ is more delicate to handle. Indeed, it can be
shown (See Bonazzola {\it et al.} 1990 \cite{bonam90}) that for
$F[r,\th,\ph,t]$ to be a $C^{\infty}$ function, the coefficients $F_{lm}$ must
vanish as $r^l$ at $r=0$. It means that the $F_{lm}[r,t]$ are symmetric with
respect to the inversion $r \rightarrow -r$ if $l$ is even, and anti-symmetric
in the opposite case. We shall then distinguish the two cases $l$ even and
$l$ odd.\\

{\em Case $l > 1$ even}\\
\samepage
The functions $H^{l}_n[r]=T_{2n}[r]-T_{2n+2}[r] $ are even
and vanish as $r^2$ at the origin. Therefore the quantity $(\frac{\d^2}
{\d r^2}+ \frac{2}{r} \frac{\d}{\d r}-\frac{l(l+1)}{r^2})H^l_n[r]$ is
regular at the origin and is retained.\\

{\em Case $l > 1$ odd}\\
The functions $H^l_n[r]=(2n+1)T_{2n+1}[r]+(2n-1)T_{2n+3}[r]$
are anti-symmetric with respect to the inversion $r \rightarrow -r$ and vanish
as $r^3$ at $r=0$. Therefore the quantity $(\frac{{\d}^2}{{\d} r^2}+\frac{2}{r}
\frac{{\d}}{{\d} r}-\frac{l(l+1)}{r^2})H^l_n[r]$ is also regular and
completes our basis.
\end{itemize}

Note that by this way, we expand the solution with a set of functions that
satisfy minimal conditions of regularity, excepted for $l=0$ and $l=1$.
It means that they vanish as $r^2$ or $r^3$, in order to make regular any
term in the equation, instead of vanishing as $r^l$ as they should to form
a $C^\infty$ function. Until now, in all the problems that our group treated,
these minimal regularity conditions were sufficient (see
Bonazzola {\it et al.} 1999 \cite{bonagm99} for more details). But we shall
see later that they make the Euler equations unstable.\\

To end with the scalar case, we shall show how the Eq.(\ref{heatleg}) can be
written when the time discretization is performed and a $2^{nd}$ order implicit
scheme used\footnote{Due to the use of Chebyshev polynomials, the Courant
conditions for the explicit problem formulation are quite severe. In the
present case $\Delta\, t_{max} $ is proportional to $1/N_{max}^4$ (see
Gottlieb {\it et al.} 1977 \cite{gott77}), therefore implicit or semi-implicit
formulation of the problem is required.}:
\begin{eqnarray} \label{heatimp}
& F_{nlm}^{j+1}- \alf \frac{\Dt t}{2} \sum_{p=0}^{\infty}A^l_{pn}F_{plm}^{j+1}=
F_{nlm}^{j}+ {\Dt t} \l(\alf \frac{1}{2} \sum_{p=0}^{\infty}A^l_{pn}
F_{plm}^{j}+ S_{nlm}^{j+1/2}\r).
\end{eqnarray}

Here $A^l_{jn}= \langle H^l_j\,O^l\,H^l_n \rangle$ is the matrix of
the operator $O^l=\frac{{\d}^2}{{\d}r^2}+\frac{2}{r} \frac{{\d}}{{\d}r }
-\frac{l(l+1)}{r^2}$. The index $j+1/2$ means a quantity $S$ computed
at the time $t^j+\Delta t/2$, obtained by extrapolating this quantity
using its (known) values at time $t^{j-1}$ and $t^j$: $S^{j+1/2}=
\l(3 S^j-S^{j-1}\r)/2 $. The left hand side operator can be easily reduced to a
pentadiagonal one, and then inverted with a number of arithmetic operations
that is proportional to the maximum value $N_{max}$ of $N$.\\

In the case of a space dependent heat conductivity $\alpha [r,\th,\ph]$,
we can use a semi-implicit method (see Gottlieb {\it et al.}
1977\cite{gott77}). It consists in solving the equation
\begin{eqnarray} \label{heatsim}
&F_{nlm}^{j+1} - (a + b r^2) \frac{\Dt t}{2} \sum_{p=0}^{\infty}A^l_{pn}
F_{plm}^{j+1}=\\
&F_{nlm}^{j} + \frac{\Dt t}{2} (a + b r^2) \sum_{p=0}^{\infty}A^l_{pn}
F_{plm}^{j}+ {\Dt t} \l( S_{nlm}^{j+1/2} + \l(\alf - a- b r^2 \r)
{\l(\sum_{p=0}^{\infty}A^l_{pn}F_{plm}\r)}^{j+1/2} \r) \nonumber
\end{eqnarray}
where $a$ and $b$ are two coefficients that satisfy the condition $a+br^2
\ge \alpha$ everywhere and that we choose in such a way that $a + b r^2$
takes the same values as $\alf$ at the boundaries.\\

The new matrix in the L.H.S. is again a pentadiagonal matrix. Boundary
conditions are imposed by adding a homogeneous solution of the
Eq.(\ref{heatimp}) or of the Eq.(\ref{heatsim})\footnote{The case
$\alpha[1,\th,\ph] =0$ is called degenerate. In this case no BC are allowed,
and $a+b$ must be $=0$.}. The reader can find in the quoted literature more
details on the spectral methods, and on how to implement boundary conditions.\\

To conclude, note that in the present example, we expand the solution in
spherical harmonics. But it turns out that in general it is
more convenient to use linear combinations of Chebyshev polynomials to treat
the expansion in $\th$. The philosophy of the spectral methods then consists
in performing simple operations such as computing derivatives, primitives,
integrals, multiplications or divisions by $r$, $r^2$, $\sin \th$ or $\cos \th$
in the coefficients space, and by making multiplications of functions in the
configuration space. If necessary, the passage to a Legendre representation is
performed with a matrix multiplication.

\subsection{Analysis in vectorial case} \label{vectorial}
\subsubsection{Stability of numerical vectorial PDE} \label{instab}

We have seen how to handle coordinates singularities appearing in scalar PDE
when spherical coordinates are used. But when treating vectorial PDE, as EE or
NSE, the singularities are more malicious: a single look at the components
of NSE [Eq.(\ref{NSEDC})] should be sufficient to convince the reader. The
singular terms cannot be handled only by laying down analytical properties, as
it was done in the above scalar equation. Indeed, the dependence existing among
the different spherical components must be taken into account in order to make
singular terms compensate each others. A simple example will illustrate this
crucial point.\\

Consider the following constant divergence free vector $\vec{V}$ of Cartesian
components: $V_x=0,\;V_y=0,\; V_z=1$. Its spherical components are
$V_r=-\cos \th; V_{\th}=\sin \th; V_\ph=0$. Then, we have ${\dv}
\vec{V}=\frac{\pt V_r}{\pt r}+ \frac{2}{r} V_r + \frac{1}{r}(\frac{\pt
V_{\th}}{\pt \th} +\frac{\cos \th} {\sin \th} V_{\th})=0$. A small error (for
example a round-off error) in the computation of components will obviously
forbid an exact compensation between the two singular terms $2V_r/r$ and
$\frac{1}{r}(\pt_\th V_{\th} + \frac{\cos \th} {\sin \th} V_{\th})$. The
consequence is the creation of high order coefficients and possible
instabilities appearing in the iterative process.\\

Indeed, we found that the hydro-code for solving linearized EE was unstable.
As expected, high frequency terms were exploding after few hundred time-steps
(few periods of the {\it r}-mode) either in the case of the anelastic
approximation or in the case of the incompressible approximation
(${\dv}\vec W =0$). As explained above, the main reason of this instability
was the non exact compensation of the singular terms contained in the
different source terms in the Poisson equations (\ref{poisdiv})
and (\ref{anent}) (see below).\\

To overcome this problem, we define two angular potentials $T_h$ and
$P_o$ in a such a way that
\begin{eqnarray} \label{potan}
W_\th=\pt_{\th} \, P_o-\frac{1}{\sin \th} \pt_{\ph}\,T_h \\
W_\ph =\frac{1}{\sin \th} \pt_{\ph} \, P_o+\pt_{\th}\,T_h. \nonumber
\end{eqnarray}  

If $r\,P_o$ and $T_h$ are any arbitrary set of regular functions, we are now
sure that the corresponding components $W_\th$ and $W_\ph$ will make the
divergent terms appearing in the vectorial PDE compensate each other. Moreover,
the system of equations (\ref{potan}) can be easily inverted. First, taking the
angular divergence of $W_\th$ and $W\ph$: \mbox{${\dv}_{\th \ph} \vec{W}
\widehat{=} (\pt_\th+\cot \th \,)W_\th +\frac{1}{\sin \th} \pt_{\ph}\, W_\ph $}
gives \mbox{$\Dt_{\th \ph} P_o ={\dv}_{\th \ph} \vec{W}$} where
\mbox{$\Dt_{\th \ph}\widehat{=} \pt^2_\th+\cot \th \pt_\th +\frac{1}
{\sin^2 \th} \pt^2_\ph $}. If \mbox{${\dv}_{\th \ph} \vec{W}$} is then expanded
in spherical harmonics, we immediately have
\be \label{P_o} P_{o,lm}=-\frac{1}{l(l+1)} ({\dv}_{\th \ph}\vec{W})_{lm}.
\ee

It is the same to compute $T_h$. Taking the angular curl of $\vec{W}$,
we obtain \be \label{T_h} \Dt_{\th \ph} T_h= \rm{curl}_{\th \ph} \vec{W} \ee
where $\rm{curl}_{\th \ph} \vec{W} \widehat{=} -\frac{1}{\sin \th}
(\pt_\ph W_\th -\pt_\th (\sin \th \,W_\ph))$.\\

To complete this procedure, we have to guarantee that $W_r$ can also
compensate the singular terms generated by the components $W_\th $ and
$W_\ph$. A satisfactory way to proceed is to take as a new variable
the quantity ${\dv} \vec{W}$. Indeed, once the quantity
$g[r,\th,\ph]={\dv} \vec{W} $ is known, it turns to be easy to find
$W_r$. Bearing in mind that \mbox{${\dv} \vec{W}=\pt_r W_r +2\,W_r/r
+1/r^2\,{\dv}_{\th \ph} \vec{W}$}, we have, after an expansion in
spherical harmonics\footnote{It is important to take into account the
following properties of the spherical harmonics decomposition of the
functions $P_o$ and $T_h$: for a given $l$ the product by $r$ of the
component $W_{r,lm} $ of $\vec{W}$, a regular vectorial function,
vanishes as $r^{l-1}$. The corresponding poloidal part $P_{o,lm} $
behaves in the same way.  The associated toroidal part is
$T_{o,(l-1)m}$ and is therefore a regular function.}
\be \label{W_r}
W_{r,lm}=\frac{1}{r^2} \int_0^r u\,\l( u\, {g [u,\th,\ph]}_{lm} +
l(l+1)\,P_{o,lm} [u,\th,\ph] \r) {\d}u.
\ee

Thus, at each time step, the strategy consists in calculating the potentials
$P_o$ and $T_h$, and then in calculating back the components $W_\th$ and
$W_\ph$. The  component $W_r$ is also computed at each time step by using the
Eq.(\ref{W_r}).\\ 

Finally, note that with this procedure, the potentials $P_o$, $T_h$ and the
component $W_r$ have the correct analytical behaviour on the axis $e_z$, i.e
they vanish as $\sin^m \th$. Thus, the instability of the EE hydro-code
was drastically reduced, but not completely suppressed. It still kept on
growing, but about ten times slower than before. The reason of this residual
instability was that, at each time step, the round-off errors did not have the
correct analytical properties at $r=0$. Indeed, they did not vanish as $r^l$.
Consequently high spatial frequency terms were still generated that did not
have the good analytical properties and a runaway toward the instability was
once again generated. To avoid this phenomenon, we project, at each time step,
all the scalar quantities $({\dv} \vec{W})_{lm}[r]$, $P_o,_{lm}[r]$ and
$T_h,_{lm}[r]$ on a Legendre space $P^l_n[r]$ in such a way to satisfy the
analytical conditions. Note that the necessity of this
procedure is due to the fact that in the EE, solved with spectral methods,
no dissipative term (numerical or physical) is present. Consequently the
numerical instability are not dumped.

\subsubsection{Solving the vectorial heat equation} 

Now, more details concerning the algorithm to solve vectorial PDE with spectral
methods. In solving NSE, we need to solve a vectorial heat equation that can
be reduced to an equation of the type
\be \label{Vecheat}
\frac{\pt \vec{\hat{B}}}{\pt t}+\nb \wedge(\mu \nb \wedge \vec{\hat{B}})
=\vec{\hat J}
\ee
where $\vec{\hat J}$ is the divergence free term of the source. We remember
that it is obtained by the decomposition of the source term $\vec{J}$
in a potential part $\vec{\nb} \phi$ and a divergence free part
$\vec{\hat J}$: $\vec{J}=\vec{ \hat J} + \vec{\nb} \phi$.\\

The Eq.(\ref{Vecheat}) can be written in the following form:
\be \label{Vecheam}
\frac{\partial \vec{B}}{\partial t}=
\mu\,\triangle \vec{B}+(\nb  \wedge \vec{B})
\wedge \vec{\nb} \mu+ \vec{\hat J}.
\ee

From a numerical point of view, the term $\vec{S}=(\nb \wedge \vec{\hat B})
\wedge \vec{\nb} \mu$ is considered as a source and is computed by a second
order scheme using its values at the times $t_{j-1}$ and $t_{j-2}$.\\

Therefore, we have to solve the equation
\be \label{Vecheaf}
\frac{\pt \vec{B}}{\pt t}=
\mu \Delta \vec{B} + \vec{\hat J}+\vec{S}
\ee
with the condition $\vnb \cdot \vec{\hat B}=0$.\\

The technique generally used was already described in Bonazzola {\it et al.}
1999 \cite{bonagm99}. Nevertheless, we adopted here a slightly different
approach which gives more accurate results.\\

To solve the Eq. (\ref{Vecheaf})
it is convenient to introduce the poloidal and toroidal potentials $P_o$
and $T_h$ as defined in appendix \ref{instab}. Here we will only consider the
case of $\mu=1$, the semi-implicit scheme which must be used in the more
general case being straightforward. With a second order scheme, we have in the
harmonic representation
\begin{eqnarray} \label{heatfd}
&P^{j+1}_{o,lm}-\frac{1}{2} \Dt t\l( \frac{{\d}^2 P^{j+1}_{o,lm}}
{{\d}r^2}+\frac{2}{r} \frac{{\d} P^{j}_{o,lm}}{{\d}r}
-\frac{l(l+1)}{r^2}P^{j+1}_{o,lm} -\frac{2}{r^2} \hat{B_r}_{lm}^{j+1} \r)
= \\
& P^j_{0,lm}+\Dt t
\l( \frac{1}{2}\l(\frac{{\d}^2 P^j_{o,lm}}{{\d}r^2}+\frac{2}{r}
 \frac{{\d} P^j_{o,lm}}{{\d}r}-\frac{l(l+1)}{r^2}P^j_{o,lm}
-\frac{2}{r^2} \hat{B_r}_{lm}^j\r)+\Pi^{j+1/2}_{lm} \r) \nonumber
\end{eqnarray}
where $\Pi$ is the poloidal component of $\vec{\hat{J}}+\vec{S}$. Note the
presence of a singular term $\frac{2}{r^2} B^{j+1}_r$ in this equation. Bearing
in mind that ${\dv} \vec{\hat{B}}=0 $ and that $P^{j+1}_{o,lm}$ vanishes as
$r^{l-1}$, it is obvious that there should be a compensation between these two
terms. Nevertheless, we shall slightly modify the Eq.(\ref{heatfd}) in order
to obtain more easily this compensation:
\begin{eqnarray} \label{heatfdm}
& P^{j+1}_{o,lm}-\frac{1}{2} \Dt t\l( \frac{{\d}^2 P^{j+1}_{o,lm}}
{{\d}r^2}+\frac{2}{r} \frac{{\d} P^{j}_{o,lm}}{{\d}r}
-\frac{l(l-1)}{r^2}P^{j+1}_{o,lm} \r)
= \\
& P^j_{0,lm}+\Dt t \l( \frac{1}{2}\l(\frac{{\d}^2 P^j_{o,lm}}{{\d}r^2}+
\frac{2}{r}\frac{{\d} P^j_{o,lm}}{{\d}r}-\frac{l(l-1)}{r^2} P^j_{o,lm}\r)
- \frac{1}{r^2}\l(\hat{B_r}^{j}_{lm}+l\,P_{0,lm}^{j}+\hat{B_r}^{j+1}_{lm}+l\,
P_{0,lm}^{j+1}\r)+\Pi^{j+1/2}_{lm} \r).
\nonumber
\end{eqnarray}
Once $P_{o,lm}^{j+1}$ known, $\hat{B}_{r,lm}^{j+1}$ is obtained as explained
above [see Eq.(\ref{W_r}), with $g=0$].\\

Finally, the equation for the toroidal part of $\vec{\hat{B}}$ reduces
to an ordinary scalar heat equation: 
\begin{eqnarray} \label{heattdm}
T^{j+1}_{h,(l-1)m}+\frac{1}{2} \Dt t\l( \frac{{\d}^2 T^{j+1}_{h,(l-1)m}}
{{\d}r^2}+\frac{2}{r} \frac{{\d} T^{j}_{h,(l-1)m}}{{\d} r}
-\frac{l(l-1)}{r^2}T^{j+1}_{h,(l-1)m} \r)
=     \\
T^j_{h,(l-1)m}+\Dt t
\l( \frac{1}{2}\l(\frac{{\d}^2 T^j_{h,lm}}{{\d}r^2}+\frac{2}{r}
 \frac{{\d} T^j_{h,lm}}{{\d} r}-\frac{l(l-1)}{r^2} T^j_{h,lm}\r) 
+\th^{j+1/2}_{(l-1)m} \r) \nonumber
\end{eqnarray}
where $\th$ is the toroidal component of $\vec{\hat{J}}+\vec{S}$.

\section{Solving Euler or Navier-Stokes equations} \label{implem}

In this section, we shall show how to implement the different schemes proposed
in the Section \ref{approx} to handle the different time scales appearing in
the EE for a slowly rotating star. In all cases, for simplicity, we will only
consider the linearized equations. First, we will begin with the case where
the exact system of equations is to solve. Then we shall consider the
implementation of physical approximations and finally the problem of BC.

\subsection{Exact Euler equations} \label{exsol}

As usual, we start by doing a decomposition of the velocity vector $\vec{W}$
in a divergence free component $\vec{\hat{W}}$ and a potential one:
\be \label{Decomp}
\vec{W}=\vec{\hat{W}} + \vec{\nb \Psi}, \; \; {\dv} \vec{\hat{W}}=0.
\ee

In a second order time scheme, the divergence free component of the EE is then
written as
\be \label{dflEE}
 \vec{\hat{W}}^{j+1}=\vec{\hat{W}}^j+\Dt t\,
\lbrack -2\vec{\Omega} \wedge \vec{W}+ \vec{F} \rbrack^{j+1/2}_{DF}
\ee

while the potential part is
\be \label{plEE}
\Psi^{j+1}=\Psi^j-\Dt t\,\lbrack
\ph_{\Omega}+ h +\ph \rbrack ^{j+1/2}.
\ee

In the Eq.(\ref{dflEE}), the subscript $DF$ means the divergence free
component of the external force. In the Eq.(\ref{plEE}), the subscript
$\Omega$ is for the potential component of the Coriolis force $2 \vec{\Omega}
\wedge \vec{W} $, $\ph$ is the potential component of some other exterior
force and $h$ is the variation of the enthalpy. Moreover, we also have the
baryonic number (or mass) conservation equation:
\be \label{mascon}
h^{j+1}=h^j-\Dt t \l( \l(\hat{W}_r +
\frac{\pt \Psi}{\pt r}\r) \, \frac{\pt H_0}{\pt r}
+ \Gamma H_0 \Dt \Psi \r) ^{j+1/2}
\ee
where $H_0$ is the enthalpy of the non perturbed configuration. In this
explicit second order scheme, the required typical value of the time step
$\Dt t$, which guarantees the numerical stability, is determined by the term
$\pt_r H_0$ that is much larger than the Coriolis term for a slowly rotating
NS.\\

The implicit version of the system of equations (\ref{plEE}), (\ref{mascon})
is the following:
\begin{eqnarray} \label{implic}
&\Psi^{j+1}=\Psi^j-\Dt t\,\l(\frac{1}{2}\l(h^{j+1}+h^j\r)+ \lbrack
\ph_{\Omega}+ h +\ph \rbrack ^{j+1/2}\r) \\
&h^{j+1}= h^j-\frac{1}{2}\Dt t \l( \l( \pt_r \Psi^{j+1}
+\pt_r \Psi^j +\hat{W}_r^{j+1} +\hat{W}_r^j\r) \pt_r H_0
+ \Gamma H_0 \l(\Dt \Psi^{j+1}+\Dt \Psi^j \r) \r) \nonumber
\end{eqnarray}
where $\hat{W}_r^{j+1}$ is obtained from the Eq.(\ref{dflEE}). Solving the
above system of equations reduces to invert, for each value of $l$ and $m$ an
enneadiagonal ($2\,N_r \times 9$) matrix (9 diagonals) where $N_r$ is the
number of degrees of freedom in $r$. The generalization to the NSE is quite
obvious and we shall not give more details about it.\\

\subsection{Implementation of physical approximations} \label{physap}

As it was already explained in the Section \ref{approx}, for solving the EE or
NSE, we chose to use approximations in order to better control the results.

\subsubsection{The divergence free approximation}

The spirit of this approximation consists in replacing the mass conservation
equation (\ref{mascon}) by the condition
\be \label{frediv}
{\dv} \vec{W}=0
\ee
or equivalently $\Psi=0$. The problem is then to find the enthalpy $h$ in a
such a way that the condition (\ref{frediv}) is satisfied. This can be done
easily by solving the Poisson equation
\be \label{poisdiv}
\Dt h - {\dv} F^{j+1/2}=0
\ee
reached by taking the divergence of the EE. As the gradient of an harmonic
function is a divergence free vector, such a function can then be
added to a particular solution of the Eq.(\ref{poisdiv}) in order to satisfy
the boundary conditions. Once $h$ is obtained, the EE has to be
solved with $\vec{\hat{F}}=\vec{F}-\vec{\nb} h$. If viscous terms are
present, NSE case, a vectorial type equation must be solved.\\

\subsubsection{Anelastic approximation}

As it was already explained, the anelastic approximation consists in
neglecting the term $\pt_t h$ in the Eq.(\ref{mascon}). Once again, the game
consists in finding the enthalpy $h$ in a such a way that the following
equation is satisfied:
\be \label{anel}
\Gamma H_0\,{\dv} \, \vec{W} +W_r\pt_r H_0=0.
\ee

To find $h$, it is sufficient to derive with respect to time the
Eq.(\ref{anel}), and to replace $\pt_t \dv \vec{W}$ by its value reached
from the EE. It gives
\be \label{anent}
\Gamma \,H_0\,\Dt h+\pt_r H_0 \pt_r h
= \Gamma \,H_0\,{\dv}\vec{F}^{j+1/2} + F_r^{j+1/2} \pt_r H_0
\ee
where $\vec{F}$ contains all the force terms (Coriolis force included).\\

The solution of the above equation is achieved with a semi-implicit scheme
very similar to the one used to solve the Eq.(\ref{heat}). The problem is then
reduced to solve the following Equation (for simplicity, we shall consider
only the case of a polytropic EOS):
\be \label{anentim}
(\gam-1)(a+br^2) \Dt h+2br \pt_r h=
\bar{H_0} \Dt h +\pt_r \bar{H_0} \pt_r h + \dv
\vec{F}^{j+1/2}+ \pt_r H_0 F_r^{j+1/2}
\ee
where $a$ and $b$ are two constants defined as was done in solving 
the Eq.(\ref{heat}). The L.H.S. operator can be easily inverted and the
Eq.(\ref{anentim}) can be solved by iteration (see
Gourgoulhon {\it et al.} 2001 \cite{gougc01}). The iteration at each time step
being quite time consuming, a more convenient strategy (although less accurate)
consists in replacing $h$ in the R.H.S. of the Eq.(\ref{anentim})
by $h^{j+1/2}$.

\subsection{Boundary conditions} \label{BC}

It was already explained how to impose BC for divergence free case. For
the system of equations (\ref{implic}) or the Eq.(\ref{anent}), it is worth
distinguishing two cases: either the enthalpy vanishes or does not vanish at
the boundary of the integration domain. Here we will only consider the case
of EE. Indeed, for NSE we chose to take a viscosity that vanishes at the
surface to avoid the need of further BC.\\

If $H_0 \, |_{r=1} >0$, the solution is quite easy : the Eq.(\ref{anent}) or
the system of equations (\ref{implic}) admit an homogeneous solution that can
be used to satisfy one BC.\\

If $H_0(1)$ vanishes, the system of equations (\ref{implic}) or the Eq
(\ref{anent}) are degenerate and therefore no BC can be imposed. But we shall
show that, in this case, the correct BC are {\em automatically} satisfied.\\

Indeed, consider first the case of the linearized exact system of equations
(\ref{implic}). On the surface of the non perturbed star, we have
$\pt_t h+ \vec{W} \cdot \vnb H_0$ that is the the correct BC. The surface
$H_0+h=0$ then defines the profile of the perturbed star. To show that the
solution in the anelastic approximation also satisfies the BC, it is more
convenient to first examine the non-linear case. Here, $h$ is not
infinitesimal, and the Eq.(\ref{anel}) must be replaced with the equation
\be \label{aneln}
\Gamma (H_0+h)\,{\dv} \, \vec{W} +\vec{W} \cdot
\vec{\nb} (H_0+h)=0
\ee
that is satisfied if $h$ is a solution of the non linearized Eq (\ref{anent}):
\be \label{anentn}
\Gamma\,(H_0+h)\,\Dt h - {\dv} \vec{F}^{j+1/2}+(\vec{\nb} h-
\vec{F}^{j+1/2}) \cdot \vec{\nb} (H_0+h))=0.
\ee

Once again, the surface of the star is defined by $H_0+h=0$. But from the
Eq.(\ref{aneln}) we have the correct BC: $\vec{W} \cdot \vec{\nb}
(H_0+h)\,|_{H_0+h=0}=0$. The surface of the star is then an unknown
quantity that is determined by the relation $H_0+h=0$ in solving the
Eq.(\ref{anentn}) by iteration. This technique was developed in a
different context in the already quoted paper \cite{gougc01}.\\

In the linear case, we have $W_r=0 $ at $r=1$ [{\it cf.} Eq.(\ref{anel})],
and the enthalpy does not vanish on the non-perturbed surface of the star. But
once again the free surface of the star can be obtained by looking for the
position where the total enthalpy $H_0+h=0$ vanishes. This surface coincides
within first order quantities with the non perturbed surface $r=1$. Note that
if $\gam>0$ the pressure $P|_{r=1} \propto h^{\gam+1}|_{r=1}$ vanishes
within terms $o(h)$.

%
\def\aa{Astron. {\rm \&} Astrophys. }                          
\def\aapr{Astron.{\rm \&}Astrophys., Rev. }                      
\def\aaps{Astron.{\rm \&}Astrophys., Supp. }                     

\def\apj{Astrophys. J. }                         
\def\apjl{Astrophys. J., Lett. }                 
\def\apjs{Astrophys. J., Supp. }                 

\def\cqg{Class. Quant. Grav.}                   

\def\mnras{Mon. Not. of the Royal Astron. Soc. }         
        
\def\apss{Astron. and Space Sci. }               

\def\jfm{J. of Fluid Mech. }
\def\ijmpd{Int. J. of Mod. Phys. D }
\def\jas{J. Atm. Sci. }
\def\nat{Nature }                                        

\def\nphys{Nucl. Phys. }                                 
\def\nphysa{{Nucl. Phys. A }}                            
\def\nphysb{{Nucl. Phys. B }}                            

\def\ptrsla{Phil. Trans. R. Soc. Lond. A }
\def\phmag{Phil. Mag. }
\def\physrep{Phys. Rep. }                                
\def\physl{Phys. Lett. }                                 
\def\physla{Phys. Lett. A }                              
\def\physlb{Phys. Lett. B }                              

\def\physr{Phys. Rev. }                          
\def\physra{Phys. Rev. A: General Physics }              
\def\physrb{Phys. Rev. B: Solid State }          
\def\physrc{Phys. Rev. C }                               
\def\physrd{Phys. Rev. D }                               
\def\physre{Phys. Rev. E }                               
\def\physrl{Phys. Rev. Lett. }                       
\def\qjrms{Quart. J. Roy. Meteor. Soc. }
\def\sicomp{SIAM J. on Comp. }                           
\def\jcomp{J. of Comp. Phys. }             
\def\jcam{J. of Comp. and App. Math. }     
\def\ssr{Space Sci. Rev. }                       


\begin{thebibliography}{10}

\bibitem{karino01}
S. Karino, S. Yoshida and Y. Eriguchi,
\newblock \physrd {\bf 64}, 024003 (2001).

\bibitem{lockandf01}
K.H. Lockitch, N. Andersson and J.L. Friedman,
\newblock \physrd {\bf 63}, 024019 (2001).

\bibitem{anders98}
N. Andersson, \newblock \apj {\bf 502}, 708 (1998).

\bibitem{thom80}
W. Thomson, \newblock \phmag {\bf 10}, 155 (1880).

\bibitem{rieut01}
M. Rieutord, \newblock \apj {\bf 550}, 443 (2001);
Erratum - ibid.  {\bf 557}, 493 (2001).

\bibitem{chandra70}
S. Chandrasekhar, \newblock \physrl {\bf 24}, 611 (1970).

\bibitem{friedschu78}
J.L. Friedman and B.F. Schutz, \newblock \apj {\bf 221}, 937 (1978);
 {\bf 222}, 281 (1978).

\bibitem{friedmor98}
J.L. Friedman and S.M. Morsink, \newblock \apj {\bf 502}, 714 (1998).

\bibitem{bil98}
L. Bildsten, \newblock \apjl {\bf 501}, L89 (1998).

\bibitem{andko01}
N. Andersson and K.D. Kokkotas, \newblock \ijmpd {\bf 10}, 381 (2001).

\bibitem{friedlock01}
J.L. Friedman and K.H. Lockitch, \newblock review in
{\it Proceedings of the 9th Marcel Grossman Meeting},
edited by V. Gurzadyan, R. Jantzen and R. Ruffini, (World Scientific,
Singapore, 2002).

\bibitem{lintv01}
L. Lindblom, J.E. Tohline and M. Vallisneri,
\newblock \physrl {\bf 86}, 1152 (2001);
\newblock \physrd {\bf 65}, 084039 (2002).

\bibitem{arr02}
P. Arras, E.E. Flanagan, S.M. Morsink, A.K. Schenk, S.A. Teukolsky and
I. Wasserman, \newblock Submitted to \apj, 25 pages.
\newblock astro-ph/0202345.

\bibitem{lesieur87}
M. Lesieur, {\it Turbulence in fluids} (Martinus Nijhoff Publishers,
Dordrecht, 1987).

\bibitem{MSPS02}
S.S. Murray, P.O. Slane, F.D. Seward, S.M. Ransom and B.M. Gaensler,
\newblock \apj {\bf 568}, 226 (2002).

\bibitem{CS2002}
F. Camilo, I.H. Stairs, D.R. Lorimer, D.C. Backer, S.M. Ransom, B. Klein,
R. Wielebinski, M. Kramer, M.A. McLaughlin, Z. Arzoumanian and P. M\"uller,
\newblock  \apj {\bf 571}, L41 (2002).

\bibitem{SBJM94}
S. Bonazzola and J.A. Marck,
Annu. Rev. Nucl. Part. Sci. {\bf 44}, 655 (1994).

\bibitem{Scmi01}
G.D. Schmidt, in {\it Magnetic fields across the Hertzsprung-Russel diagram},
proceeding of the workshop, edited by G. Mathys, S.K. Solanki
and D.T. Wickramasinghe, (ASP, San Francisco, 2001).

\bibitem{scha62}
E. Schatzman,
\newblock  Ann. Astrophys. {\bf 25}, 18 (1962).

\bibitem{spr98}
H.C. Spruit,
\newblock  \aa {\bf 333}, 603 (1998).

\bibitem{spph98}
H.C. Spruit and E.S. Phinney,
\newblock  \nat {\bf 393}, 139 (1998).


\bibitem{STRO02}
T.D. Strohmayer and C.B. Markwardt,
\newblock accepted for publication in the Astrophysical Journal, 27 pages.
\newblock astro-ph/0205435

\bibitem{mar98}
F.E. Marshall, E.V. Gotthelf, W. Zhang, J. Middleditch and Q.D. Wang,
\newblock  \apjl {\bf 499}, L179 (1998).

\bibitem{haen89}
P. Haensel and J.L. Zdunik,
\newblock \nat {\bf 340}, 617 (1989).

\bibitem{koj98}
Y. Kojima,
\newblock \mnras {\bf 293}, 49 (1998).

\bibitem{ruoff01}
J. Ruoff and K.D. Kokkotas,
\newblock \mnras {\bf 328}, 678 (2001).

\bibitem{ruoff01b}
J. Ruoff and K.D. Kokkotas,
\newblock \mnras {\bf 330}, 1027 (2002).

\bibitem{koj92}
Y. Kojima,
\newblock \physrd {\bf 46}, 4289 (1992).

\bibitem{nom87}
K. Nomoto and S. Tsurita,
\newblock \apj {\bf 312}, 711 (1987).

\bibitem{yak01}
D.G. Yakovlev, A.D. Kaminker, O.Y. Gnedin and P. Haensel,
\newblock \physrep {\bf 354}, 1 (2001).

\bibitem{blan937}
L. Blanchet,
\newblock \physrd {\bf 47}, 4392 (1993); {\bf 55}, 714 (1997).

\bibitem{rezz99}
L. Rezzolla, M. Shibata, H. Asada, T.W. Baumgarte and S.L. Shapiro.
\newblock \apj {\bf 525}, 935 (1999).

\bibitem{haen01}
P. Haensel, K.P. Levenfish and D.G. Yakovlev,
\newblock \aa {\bf 357}, 1157 (2000);  {\bf 372}, 130 (2001)
and  {\bf 381}, 1080 (2002).

\bibitem{cow41}
T.G. Cowling,
\newblock \mnras {\bf 101}, 367 (1941).

\bibitem{saio82}
H. Saio,
\newblock \apj {\bf 256}, 717 (1982).

\bibitem{bonagm97}
S. Bonazzola, E. Gourgoulhon, and J-A. Marck,
\newblock \physrd {\bf 56}, 7740 (1997).

\bibitem{bonagm98}
S. Bonazzola, E. Gourgoulhon, and J-A. Marck,
\newblock \physrd {\bf 58}, 104020 (1998).

\bibitem{bonagm99}
S. Bonazzola, E. Gourgoulhon, and J-A. Marck,
\newblock \jcam {\bf 109}, 433 (1999).

\bibitem{mors53}
P.M. Morse and H. Feshbach, {\it Methods of theoretical physics}
(McGraw-Hill, New-York, 1953).

\bibitem{cut87}
C. Cutler and L. Lindblom,
\newblock \apj {\bf 314}, 234 (1987).

\bibitem{bat53}
G.K. Batchelor,
\newblock \qjrms {\bf 79}, 224 (1953).

\bibitem{oguph62}
Y. Ogura Y. and N.A. Phillips,
\newblock \jas {\bf 19}, 173 (1962).

\bibitem{lat76}
J. Latour, E.A. Spiegel, J. Toomre and J.P. Zahn,
\newblock \apj {\bf 207}, 233 (1976).

\bibitem{dinrieu01}
B. Dintrans and M. Rieutord,
\newblock \mnras {\bf 324}, 635 (2001).

\bibitem{rieudin02}
M. Rieutord and B. Dintrans,
\newblock astro-ph/0206357.

\bibitem{haen01b}
P. Haensel, in {\it Physics of NS interiors}, proceeding of the workshop,
edited by D. Blaschke, N.K. Glendenning and A. Sedrakian,
(EC$\text{T}^*$, Trento, 2000).

\bibitem{linowus00}
L. Lindblom, B.J. Owen and G. Ushomirsky,
\newblock \physrd {\bf 62}, 084030 (2000).

\bibitem{beyer99}
H.R. Beyer and K.D. Kokkotas,
\newblock \mnras {\bf 308}, 745 (1999).

\bibitem{spr99}
H.C. Spruit,
\newblock \aa {\bf 341}, L1 (1999).

\bibitem{rezls00}
L. Rezzolla, F.K. Lamb and S.L. Shapiro,
\newblock \apj {\bf 531}, L141 (2000).

\bibitem{rezlms01}
L. Rezzolla, F.K. Lamb, D. Markovic and S.L. Shapiro,
\newblock \prd {\bf 64}, 104013 (2001).

\bibitem{rezlms01b}
L. Rezzolla, F.K. Lamb, D. Markovic and S.L. Shapiro,
\newblock \prd {\bf 64}, 104014 (2001).

\bibitem{sch02}
A.K. Schenk, P. Arras, E.E. Flanagan, S.A. Teukolsky and I. Wasserman,
\newblock \prd {\bf 65}, 024001 (2002). 

\bibitem{cook96}
G.B. Cook, S.L. Shapiro and S.A. Teukolsky,
\newblock \physrd {\bf 53}, 5533 (1996).

\bibitem{hartl67}
J.B. Hartle,
\newblock \apj {\bf 150}, 1005 (1967). 

\bibitem{bona93}
S. Bonazzola, E. Gourgoulhon, M. Salgado and J-A. Marck,
\newblock \aa {\bf 278}, 421 (1993).

\bibitem{bonam90}
S. Bonazzola and J.A. Mark,
\newblock \jcomp {\bf 87}, 201 (1990).

\bibitem{gott77}
D. Gottlieb and S.A. Orszag, {\it Numerical Analysis of spectral Methods:
Theory and Applications} (Society for Industrial and Applied Mathematics,
Philadelphia, 1977).

\bibitem{gougc01}
E. Gourgoulhon, P.Grandcl\'ement, K. Taniguchi, S. Bonazzola, and J.A. Mark,
\newblock \physrd {\bf 63}, 064029 (2001).

\end{thebibliography}
\end{document}